\documentclass[aps,prd,reprint,amsmath,amssymb,superscriptaddress,floatfix,nofootinbib]{revtex4-2}

\newcommand{\msun}{\ensuremath{\mathrm{M}_\odot}}
\usepackage{amsmath}
\usepackage{subcaption}
\usepackage{hyperref}
\usepackage{orcidlink}
\usepackage{caption}

\usepackage{ragged2e}
\DeclareCaptionJustification{justified}{\justifying}
\makeatletter
\newcommand*\bibmaxauthors{3}
\newcount\bib@aucount
\newcount\bib@autotal
\newbox\bib@binA
\newbox\bib@binB
\def\bib@author@name{author}
\def\bib@au@tally#1#2{\global\advance\bib@autotal\@ne}
\def\bib@au@info#1#2{%
  \advance\bib@aucount\@ne
  \ifnum\bib@aucount>\bibmaxauthors\relax
  \else
    \bib@info@save{#1}{#2}%
    \ifnum\bib@aucount=\bibmaxauthors\relax
      ,\ \emph{et~al.},\ %
      \setbox\bib@binB\hbox\bgroup
    \fi
  \fi
}
\def\bibfield#1#2{%
  \def\bib@tempa{#1}%
  \ifx\bib@tempa\bib@author@name
    \global\bib@autotal\z@
    \begingroup \let\bibinfo\bib@au@tally \setbox\bib@binA\hbox{#2}\endgroup
    \ifnum\bib@autotal>\bibmaxauthors\relax
      \bib@aucount\z@
      \let\bib@info@save\bibinfo
      \let\bibinfo\bib@au@info
      #2%
      \egroup
      \let\bibinfo\bib@info@save
    \else
      #2%
    \fi
  \else
    #2%
  \fi
}
\makeatother

\begin{document}

\title{Normalizing Flow Emulator for Gravitational Wave Detection Probability in the LIGO/Virgo/KAGRA O4 Observing Run}

\author{Julius Gassert\,\orcidlink{0009-0008-2754-1946}}
\email{julius.gassert@campus.lmu.de}
\affiliation{McWilliams Center for Cosmology and Astrophysics, Department of Physics, Carnegie Mellon University, Pittsburgh, PA 15213, USA}
\affiliation{University Observatory, Faculty of Physics, Ludwig-Maximilians-Universit\"at M\"unchen, Scheinerstr.~1, 81679 Munich, Germany}

\author{Ignacio Maga\~na~Hernandez\,\orcidlink{0000-0003-2362-0459}}

\affiliation{McWilliams Center for Cosmology and Astrophysics, Department of Physics, Carnegie Mellon University, Pittsburgh, PA 15213, USA}

\author{Ariel J.~Amsellem\,\orcidlink{0000-0003-3433-2698}}

\affiliation{McWilliams Center for Cosmology and Astrophysics, Department of Physics, Carnegie Mellon University, Pittsburgh, PA 15213, USA}

\author{Antonella Palmese\,\orcidlink{0000-0002-6011-0530}}

\affiliation{McWilliams Center for Cosmology and Astrophysics, Department of Physics, Carnegie Mellon University, Pittsburgh, PA 15213, USA}

\author{Barnabas P\'oczos\,\orcidlink{0000-0002-2898-4705}}

\affiliation{McWilliams Center for Cosmology and Astrophysics, Department of Physics, Carnegie Mellon University, Pittsburgh, PA 15213, USA}
\affiliation{Machine Learning Department, Carnegie Mellon University, Pittsburgh, PA 15213, USA}

\date{\today}

\begin{abstract}

Efficient and accurate estimation of the detection probability of a gravitational wave event given its source parameters remains a challenging problem because it conventionally requires expensive Monte Carlo injection campaigns. We present a machine learning-based approach using Normalizing Flows (NFs) to estimate the detection probability of gravitational wave events for a LIGO/Virgo/KAGRA O4-like detector network configuration. We utilize the publicly available injection dataset as training and test data. We test our method and implementation by evaluating the raw NF performance as a generative model based on its ability to recover the correct parameter distributions of detected events. In addition, we compare the detection efficiencies predicted by the NF emulator with those from Monte Carlo importance sampling and find good agreement. Furthermore, estimating the detection efficiency for a wide variety of population models, including those with sharp features, can be more precise than estimates obtained via Monte Carlo importance sampling. We show that the NF emulator can be used in population analysis of gravitational wave sources, similar to previous neural network approaches. 
The training code and trained models from this work are publicly available.

\end{abstract}


\maketitle

\section{Introduction \label{sec:introduction}}

In virtually every field of physics, selection effects play a crucial role in designing experiments and detectors as well as the subsequent data analysis and inference process \cite{Malmquist1922, Malmquist1925, Toledo2019SelectionEffects}. The direct detection of gravitational waves (GWs) with modern interferometers such as Advanced LIGO \cite{advancedLIGO2015}, Advanced Virgo \cite{advancedVirgo2014}, and KAGRA \cite{KAGRA2021} is most sensitive to signals from ``loud'' mergers of compact objects. Therefore, these GW detectors are, within limits and generally speaking, less sensitive to more distant and less massive events. The selection effects of detected binary parameters from GW detectors depend on: a) intrinsic (masses, spins, tilt angles, etc.) and extrinsic (distance to observer, event on-sky position, relative inspiral orientation, etc.) parameters $\theta$ of a binary merger; b) detector (network) configuration; and c) detector noise \cite{Schutz1987,Finn1993,Cutler1994,Flanagan1998,Martynov2016aLIGOSensitivity}.

In recent years, the scientific community has made a significant effort to {describe and characterize} the intrinsic source population of coalescing binaries detected by LIGO-Virgo-KAGRA (LVK) (see e.g.  \cite{Abbott2019GWTC12Population,Safarzadeh2020MChieff,Tiwari2021ModelingPop,Baxter2021,Rinaldi2021GMMPop,Zevin2021Pop,GWTC2Population2021,Ruhe2022NF4POP,GWTC3Population2023,Farah2023PopulationBump,Leyde2024NPEPop,Callister2024Population,Callister2024PopTour,MaganaHernandez2025MassBump,GWTC4Population2025,Antonini2025PISNGap,Tong2026PISNGap,Ray2026NoPISNGap,GWTC5Population2026,Amsellem2026UMAPPop}). To do so, the noisy and incomplete catalogs need to be corrected for the selection function of the detectors \cite{Mandel2019SelectionBias,Vitale2021SelectionEffects}. Selection function effects also need to be corrected for in other applications such as GW cosmology \cite{Schutz1986, Pozzo2012CosmoFromGW, Abbott2017H0, SoaresSantos2019H0BrightSiren, Palmese2023H0, Alfradique2024Darksiren, IMH2024SpectralSirens, IMH2025SpectralSirens, Palmese2026GWCosmo,LIGO2026MondCosmo}. Not only in inference, but also in forward modeling (future) observations, the selection function is of vital importance \cite{
Dominik2015DetectionRates,Rodriguez2019,Belczynski2020,Bavera2020PredictingPop}, motivating the need for a fast, easy-to-use, and accurate solution for the evaluation of the selection function.

It is possible to accurately quantify selection effects \cite{Biwer2017HWInjections,Essick2025Injections} by combining generated waveform signals -- which can be computed to a very high degree of accuracy \cite{Pratten2021IMRPHenom,Essick2025Injections} -- with realistic detector noise and analyzing the data with the same search pipelines \cite{Klimenko2008CWB,Klimenko2016CWBImprovements,Adams2016MBTA,Usman2016PyCBC,Messick2017GstLAL} as in the observing campaigns. These efforts to approximate the selection function are called \textit{injection campaigns}. The resulting detections and non-detections constitute a discrete set of points in the high-dimensional parameter space, which may be used to identify a detection boundary, or rather a transition region between detectability and non-detectability. However, this transition region is not a sharp hypersurface in the parameter space but a stochastic region. Due to the stochastic noise of the detector network, pipeline search characteristics, etc., detectability is not a deterministic function of the binary parameters. The injection campaigns are very labor-intensive and require significant amounts of computational resources \cite{Essick2023SemiEstimates}, which is why several alternative approaches have been proposed. While some use analytically derivable quantities, often related to the signal-to-noise ratio (SNR) \cite{Wong2020SNREstimator, Essick2023SemiEstimates,Gerosa2024Recipes,Mould2024SNRDetection,LorenzoMedina2025Selection,Vijaykumar2026FFSemianalytic}, others have proposed machine-learning techniques to identify the boundary between detected and not detected events \cite{Gerosa2020Selection,Talbot2022,Callister2024Emulator}. In particular, \cite{Callister2024Emulator} used the injection campaign from the third observing run \cite{LIGO2023Injection} to train a fully connected neural network  emulator that classifies merger events as either detected or not detected, solely based on their event parameters {$\theta$}. The last layer of the network outputs a score $\in [0,1]$, which is interpreted as the probability of detection. \cite{Talbot2022} trained a Gaussian mixture model (GMM) to the density of detected injection data, which allows fast estimation of the detection probability given the binary parameters of a GW event. Their GMM approach is a computationally efficient alternative to the more expensive Monte Carlo resampling, mainly enabled by efficient pre-processing of the data. They test their method on the gravitational wave transient catalog (GWTC) 3 and find performance similar to that reported in other works.

These alternative approaches to estimating the detection probability allow for an accurate and tractable approach to GW population inference. This is because the selection function enters the inference and depends on the population model to be tested. They become particularly useful for population models with sharp or peaked features, where the number of samples in traditional Monte Carlo importance sampling (MC-IS) of the injection campaign data is fixed, and the effective number of samples probing these features decreases as the features become narrower. Too few effective samples increase the variance of the integral approximation. This can lead to biased likelihood estimates, limiting the search space or accuracy of traditional MC-IS in compact object population studies.

We introduce a new method for modeling the selection function in GW data analysis, using a flexible generative Normalizing Flow (NF) \cite{Tabak2013NFs,Rezende2015NFs,Kobyzev2021NFReview,Papamakarios2021NFReview} to accurately evaluate detectability. Unlike some previous approaches that treat detectability as a classification problem, our method approximates the density of GW parameters for detected GW events, $p(\theta| \mathbb{D})$, where $\mathbb{D}$ denotes a detected event. Using Bayes' theorem and the known injection population model, we can relate $p(\theta| \mathbb{D})$ to the detection probability. This density-estimation approach offers several key {features}:

\begin{enumerate}
    \item Density-ratio formulation: by applying Bayes' theorem, we can model the selection function $P(\mathbb{D} | \theta)$ via the ratio of the learned density to the injection-population density. The result is, in principle, properly normalized and does not rely on scores obtained from a classifier.
    \item Generative evaluation: like a GMM, an NF can be used as an emulator to generate synthetic detected binary merger events. The distributions of the parameters offer an informative and powerful diagnostic tool. Comparing the parameter distributions of the detected injection data with samples generated from the NF, we can assess the accuracy of the model.
    \item High-dimensional expressivity: NFs are, given the right network structure, expressive models \cite{Bogachev2005,Hung2018NAFsExpressivity,Durkan2019Flow} that can be fit to much higher-dimensional and more complex data than GMMs, while remaining computationally efficient.
    \item  No additional training data beyond the injection campaign data itself are required; the method requires only minimal preprocessing, no feature engineering, and no auxiliary loss terms, while achieving training times of $\mathcal{O}(10)$ minutes with GPU support.
\end{enumerate}

We apply our proposed method to the latest injection campaign for the O4a and O4b observing runs \cite{Essick2025Injections}. We validate our methodology by performing three main tests: First, we ensure that the NF as a generative model reproduces the same parameter distributions as in the injection data. Second, we draw new merger events according to the injection population model, evaluate the detection probability with our density-based emulator and probabilistically accept them as emulator detections. The associated parameter distributions are again compared to the true injection distributions, providing a test of the full emulator pipeline. Finally, we ensure that the detection efficiency, an important quantity in GW population studies, is correctly estimated not only for the injection population model but also for additional models. We publish the fully trained networks for O4a and O4b in an installable Python package that allows for quick and easy evaluation of the detection probability given some event parameters. The published code includes all necessary scripts to retrain the NF for future injection campaigns.

This paper is structured as follows: We first cover the data required to train and evaluate the NFs in Section \ref{sec:data} and in Section \ref{sec:methodology} we discuss the methodology in detail. {Section \ref{sec:evalmetrics} covers the two-sample tests to evaluate the emulator performance. We show the evaluation results of our emulator in Section \ref{sec:results}. Finally, we discuss our emulator and conclude} in Section \ref{sec:conclusion}.

\section{Data \label{sec:data}} 

\subsection{Injection Campaign \label{subsec:injcamp}} 

For the NF to learn the parameter space of detected events, we use the publicly available search sensitivity estimates from \cite{LIGO2025InjectionsZenodo,LIGO2026InjectionsZenodo}. These were produced by simulating compact-binary coalescence signals sent through the detector network, and by recording the recovered parameters and search statistics. Throughout this work, the term \textit{injection} refers to these simulated search sensitivity estimates. A more detailed description of the injection campaign and the associated data products can be obtained from  \cite{Essick2025Injections}.

In short, the injection campaign encompasses a few steps, beginning with estimating the noise floor of the GW detectors. Next, a set of source parameters $\left\{\theta_1, \theta_2, \dots, \theta_n\right\}$ is drawn from a population model described by its model parameters $\Lambda_\mathrm{inj}$. Combined with a random event time uniformly placed over the observing run, these parameters define the full signal for each event. Together with noise, the projected strain then makes up the recorded data, which is subsequently analyzed by search pipelines. These output additional metrics in the form of ranking statistics, including the false alarm rate (FAR). An event is usually considered detected if its FAR falls below some threshold. A commonly adopted detection threshold is $\mathrm{FAR}_\mathrm{thresh} = 1/\mathrm{yr}$ for any of the search pipelines \cite{LIGO2025GWTC4Overview}.

For our method, the population model used to generate the binary parameters of the injection data needs to be known. The primary mass is distributed according to a five-piece broken power law. The secondary masses follow a single power law conditioned on being smaller than the primary mass. Spin magnitudes follow a normal distribution and spin tilts are combined as a mixture of isotropic and uniform distributions. Redshifts are distributed uniformly in comoving volume. We define the probability density functions of the injection model in Appendix \ref{subsec:appendixB_pdfs}, and refer to \cite{Essick2025Injections} for full details. Whenever we mention the \textit{injection (population) model}, that description of how binary merger parameters are drawn is assumed. $\Lambda_\mathrm{inj}$ denotes all the population distribution parameters (often referred to as hyperparameters). Note that we omit the time and phase of the binary coalescence throughout this work. Assuming uniform distributions for each, subsequent quantities should therefore be understood as time- and phase-averaged.

The final O4a injection dataset contains around 1.5 million events, of which around one third are detected, with similar numbers in O4b. Note that originally approximately 433 million sets of event parameters were drawn in O4a, but only a fraction of those were passed on to the search pipelines since many did not satisfy the cut on the optimal network SNR of $\rho_\mathrm{opt, net} > 5$ and were therefore immediately considered non-detections.  

These injections cover the search sensitivity estimates for O4a spanning May 2023 to January 2024 and O4b spanning April 2024 to January 2025 \cite{Essick2025Injections,LIGO2025GWTC4Overview}. An injection campaign for O4c has been announced for release alongside a future update of the GWTC, motivating the development of an easy-to-use Python package that may be used for retraining and adding those injection campaigns at a later time.

\section{Methodology \label{sec:methodology}} 

The goal is to estimate the detection probability $P(\mathbb{D}| \theta)$ of a GW event, given some of its intrinsic and extrinsic parameters $\theta$. We use a machine-learning approach that requires only an injection data set as training, validation, and test data, as well as knowledge of how the injection data were drawn, particularly the injection distribution $p(\theta|\Lambda_\mathrm{inj})$. Once trained, this emulator can be used to estimate the detection probability $P(\mathbb{D}| \theta)$.

Unlike a classification-based emulator (e.g. as in \cite{Callister2024Emulator}), which optimizes a binary cross-entropy loss to identify a boundary between detection and non-detection, our approach treats the problem through density estimation. At the core of our method is an NF that models the distribution of parameters of detected events, $p(\theta | \mathbb{D})$, which is related to the detection probability via Bayes' theorem:
\begin{align} \label{eq:pdetbayes}
    P(\mathbb{D}| \theta) = \frac{p(\theta | \mathbb{D}) P(\mathbb{D})}{p(\theta)}
\end{align}

where the denominator is the prior population model of the parameters in the injection campaign, discussed in section \ref{subsec:injcamp}, $p(\theta) = p(\theta | \Lambda_\mathrm{inj})$.
The probability that an event is detected in the injection data, $P(\mathbb{D}) = P(\mathbb{D} | \Lambda_\mathrm{inj})$, is the detection efficiency (denoted by $\xi$, and further described in section \ref{subsec:deteff}). 

The choice of an NF as a density estimator for $p(\theta | \mathbb{D})$ over simpler models such as a GMM, is motivated by the 13-dimensional parameter space and complex non-linear binary merger parameter relations. GMMs scale poorly in higher dimensions and are not as expressive as NFs. GMMs need a rapidly growing number of components to capture curved or correlated manifolds, each with an $\mathcal{O}(d^2)$ covariance. In contrast, NFs act as universal density approximators much more efficiently \cite{Bogachev2005,Hung2018NAFsExpressivity,Durkan2019Flow} for continuous distributions: due to the composition of bijective transformations, they are able to effectively map any complex target distribution into a simple latent space representation, provided enough training data and model capacity.

\subsection{Detection Efficiency\label{subsec:deteff}} 

 One of the terms in eq. \ref{eq:pdetbayes} is the detection efficiency, which is evaluated as \cite{Essick2025Injections}:
\begin{align} \label{eq:deteff}
    \xi(\Lambda_\mathrm{inj}) & \equiv P(\mathbb{D} | \Lambda_\mathrm{inj}) = \int P(\mathbb{D} | \theta, \Lambda_\mathrm{inj}) p(\theta | \Lambda_\mathrm{inj}) d\theta \\ \nonumber
    &\approx \frac{N_\mathbb{D}}{N_\mathrm{gen}}
\end{align}
where $P(\mathbb{D} | \theta, \Lambda_\mathrm{inj})$ is a binary detection label. Due to the way the injection data were generated in O4, ``injection weights'' $w_i$ corresponding to the monthly PSD variation (leading to unequal acceptance fractions of GW events) need to be taken into account, so eq. \ref{eq:deteff} is estimated as \cite{Essick2021MixtureModels,Essick2023SemiEstimates}:

\begin{equation}
    \xi(\Lambda_\mathrm{inj}) \approx \frac{1}{N_\mathrm{gen}}\sum_i^{N_\mathbb{D}}w_i
\end{equation}
where $N_\mathbb{D}$ denotes all detected events, while $N_\mathrm{gen}$ refers to all binary parameter sets that were generated.

If one is interested in the detection efficiency for a population model with hyperparametyers $\Lambda_\mathrm{new}$ different from $\Lambda_\mathrm{inj}$, it is not necessary to repeat all the steps in the injection campaign. Instead, detected injections may be used as a proposal distribution for importance sampling. The integral in eq. \ref{eq:deteff} is evaluated with the injection draw distribution $p(\theta|\Lambda_\mathrm{inj})$ as the proposal, and each detection is reweighted by the ratio of the new population density to the injection draw density. This leads to an expression including the weights from the injection data ($w_i$) and the importance sampling weights $w_{i,\mathrm{model}}$ entering as a fraction of the new and injection population model densities:
\begin{align} \label{eq:deteff-MCIS}
    \hat{\xi}(\Lambda_\mathrm{new} | \Lambda_\mathrm{inj}) & \approx \frac{1}{N_\mathrm{gen}}\sum_i^{N_\mathbb{D}} w_iw_{i,\mathrm{model}}\ \mathrm{, with} \\  w_{i,\mathrm{model}} &= \frac{p(\theta_i|\Lambda_\mathrm{new})}{p(\theta_i | \Lambda_\mathrm{inj})} 
\end{align}

We will use eq. \ref{eq:deteff-MCIS} in section \ref{subsec:nffoundeval} to confirm that the emulator correctly predicts the detection efficiency for a variety of population models. This is relevant for population studies because the detector selection function term is traditionally incorporated into the Bayesian population analysis framework via the detection efficiency. However, a major limitation in population analysis using this approach is the uncertainty associated with the MC-IS integral approximation, which can lead to biased population inference results 
\cite{Mandel2019SelectionBias}.

When considering a population model $\Lambda_\mathrm{new} \neq \Lambda_\mathrm{inj}$, the original samples still trace the injection population. However, MC-IS uses those original samples to estimate the integral under the new target population. If the new target population assigns considerable weight to regions in which the original samples are sparsely distributed, for example via a Gaussian peak, a smaller number of samples will contribute substantially to the reweighted integral compared to the case of the original distribution. Therefore, the effective number of samples that inform the reweighted MC-IS integral is reduced, specifically due to regions where the new population model density distributions differ from the original draw densities. The number of effective samples, $N_\mathrm{eff}$, quantifies how many samples inform the MC-IS integral approximation and is defined as \cite{Essick2023SemiEstimates}: 
\begin{align} \label{eq:Neff}
    N_\mathrm{eff} = \frac{\left(\sum_i w_i \cdot w_{i,\mathrm{model}}\right)^2}{\sum_i w_i^2 \cdot w_{i,\mathrm{model}}^2}   
\end{align}
It has been shown that $N_\mathrm{eff}$ needs to grow faster than linearly with catalog size for population inference to remain unbiased, with a commonly used \emph{minimum} threshold of $N_\mathrm{eff} > 4N_\mathrm{obs}$ \cite{Farr2019, Essick2022Neff, Talbot2023, Heinzel2025NeffSelection}.

In addition to the standard MC-IS error of the detection efficiency, we also use the approach laid out in \cite{Callister2024Emulator} to estimate the variance of the detection efficiency estimate. Essentially, the detection of each injection is treated as a Bernoulli trial. Thus, each injection is independently detected with probability $\xi(\Lambda_\mathrm{inj})$. This implies that the number of detections is $N_\mathbb{D} \sim \mathrm{Binomial}(N_\mathrm{gen}, \xi)$. The unbiased estimator $\hat{\xi}=N_\mathbb{D}/{N_\mathrm{gen}}$ then has variance $\hat{\xi}(1-\hat{\xi})/N_\mathrm{gen}$. In the case of MC-IS, $N_\mathrm{gen}$ is replaced with $N_\mathrm{eff}$ in the variance estimate and becomes 

\begin{align} \label{eq:xi_Neff_err}
    \sigma^2(\hat{\xi}) = \hat{\xi}(1-\hat{\xi})/N_\mathrm{eff}.
\end{align}
We treat this variance estimate as a conservative bound, since the weight-based Monte Carlo variance estimate of the detection efficiency is dominated by the few largest weights and is thus poorly constrained when $N_\mathrm{eff}$ is small. 

\subsection{Probability Density of Parameters from Detected Events \label{subsec:densitydeteventparams}}

While the terms $P(\mathbb{D})$ and $p(\theta)$ in eq. \ref{eq:pdetbayes} can be calculated or at least estimated analytically, the probability density of the parameters given detection, $p(\theta | \mathbb{D})$, is not as easily available. Instead, the injection data provide point estimates of this density only. Therefore, a natural way of approximating this density is to leverage NFs. These models not only allow easy generation of synthetic data resembling the input but also provide a fast and exact evaluation of the probability density associated with an input. In this case, the inputs are the parameters $\theta$ of a GW event.

We train an NF on the detected events {in the injection data}, which can then be evaluated as $p_\mathrm{NF}(\theta) = \hat{p}(\theta | \mathbb{D})$. We denote estimated probability distributions with the $\text{`` }\hat{}\text{ ''}$ symbol. One can think of the NF as a smooth density model constrained by the support points produced via the injection data in a high-dimensional space. It should give a density $p(\theta | \mathbb{D}) \approx 0$ in the high-dimensional parameter space where no events are detected and a relatively high density in regions with many detections.

Because the NF approximates the density of the underlying parameters using a finite set of support points, the resulting estimate is inherently imperfect. Consequently, this approach may result in predicted detection probabilities $P(\mathbb{D} | \theta) > 1$, which we refer to as ``over-predictions''. Over-predictions are trimmed to a maximum value of $P(\mathbb{D} | \theta) = 1$.

The parameters $\theta$ that we have used for training and evaluation are the same as those described in section \ref{subsec:injcamp}. If the user does not provide all 13 merger parameters (see Table \ref{tab:injection_priors} for definitions), the code automatically draws the remaining parameters according to the O4 injection model, $\Lambda_\mathrm{inj}$, described in section \ref{subsec:injcamp}.

\subsection{Normalizing Flows \label{subsec:NFbasics}}

NFs \cite{Tabak2013NFs,Rezende2015NFs} are a class of generative models that perform density estimation by learning a bijective transformation from a simple base distribution to the target data distribution $p_X(x)$. Because the transformation is invertible, the data density $p_X(x)$ can be computed exactly via the change-of-variables formula, using the base density and the Jacobian determinant of the transformation. Given an invertible mapping $f: \mathbb{R}^d \rightarrow \mathbb{R}^d$ that transforms $X = f(Z)$, we write:

\begin{align} \label{eq:ChangeOfVar}
    p_X(x) = p_Z(f^{-1}(x))\left|\det \left(\frac{\partial f^{-1}(x)}{\partial x}\right)\right|
\end{align}

The task is to obtain a mapping $f$ that can transform between the spaces $X$ and $Z$. In practice, we only have a dataset $\mathcal{D} = \left\{x^{(1)},x^{(2)},\dots, x^{(N)}\right\}$ of $N$ data points $x^{(i)} \in \mathbb{R}^d$ that represent some target distribution. By parameterizing $f \rightarrow f_\varphi$ with network parameters $\varphi$, we can learn these mappings. This is done by maximizing the likelihood of the data under the model, i.e. choosing $\varphi$ of the NF to maximize $\prod_{i=1}^N \hat{p}_X(x^{(i)}|\varphi)$, where $\hat{p}_X(\cdot|\varphi)$ denotes the density obtained by inserting $f_\varphi$ into eq. \ref{eq:ChangeOfVar}. The base density $p_Z$ we use is a normal distribution, $p_Z = \mathcal{N}(0,1)$. For numerical stability, the mean negative log-likelihood over the dataset $\mathcal{D}$ is minimized:

\begin{align} \label{eq:NLL}
    \log\mathcal{L}(\mathcal{D};\varphi) &= -\frac{1}{N}\sum_{i=1}^{N} \log \hat{p}_X(x^{(i)}) \nonumber \\
    &= -\frac{1}{N}\sum_{i=1}^{N} \Big[ \log p_Z\big(f_\varphi^{-1}(x^{(i)})\big) \nonumber \\
    &\qquad + \log \left| \det \left( \frac{\partial f_\varphi^{-1}(x)}{\partial x}\Big|_{x = x^{(i)}} \right) \right| \Big]
\end{align}

A single mapping $f_\varphi$ may not be powerful enough to capture the potentially complex warping of the base distribution into the target distribution, so many mappings are sequentially chained together as layers to increase the expressivity of NFs and model complex data distributions.

A central NF design question is how to construct a family of flexible mappings that are easily inverted, with differentiable Jacobians that are fast to compute. Many network structures have been suggested in the literature (see e.g. \cite{Kobyzev2021NFReview, Papamakarios2021NFReview} for reviews on NFs), but we only cover the network type used in this work, a triangular spline flow \cite{ward2026flowjax}. Each layer combines two transformations. First, a monotonic rational-quadratic (RQ) spline \cite{Durkan2019Flow} is applied to each of the $d$ dimensions independently. For a given dimension, a bounded interval is divided into bins, with a separate RQ spline in each. The bin boundaries are called ``knots'', and their positions together with the derivatives at them parametrize the transformation. Thus, the number of knots directly controls the expressivity. A $\tanh$ transformation compresses the input into the spline's finite interval and maps it back afterwards. Since the splines act on each dimension independently, the Jacobian of this step is diagonal. However, it lacks the ability to introduce dependencies between the dimensions. These are instead created by a second transformation, a learned lower-triangular affine map that mixes the $d$ dimensions. Its Jacobian is triangular by construction. The composition of diagonal and triangular Jacobians remains triangular and the log-determinant in eq. \ref{eq:NLL} reduces to a sum, lowering its cost from $\mathcal{O}(d^3)$ to $\mathcal{O}(d)$. Consecutive layers are interleaved by random permutations of the dimensions, so that no ordering is privileged.

 General ML limitations apply to NFs as well: fixed training dataset size, noise in the training data, limited training time and finite model capacity in practice contribute to imperfect fits of the true underlying density. Scalability of NFs to high dimensions can deteriorate, but we are working in a modest regime that is not affected.

\subsection{Training the NF \label{subsec:trainNF}}

Standard ML techniques are used for training the NF. Upon loading the injection dataset and including only detected events, we are left with nearly 500,000 13-dimensional parameter combinations. Explicitly, we have a dataset $\mathcal{D} = \{x^{(1)}, \dots, x^{(N)}\}$ with $N=476,406$ for O4a and each $x^{(i)} \in \mathbb{R}^{d=13}$ representing a detected binary merger. We split the data into a training and test set at a split ratio of 0.9. The training set is split again at a split ratio of 0.8 to additionally obtain a validation set. To improve the convergence of the training, we standardize the data, i.e. subtract the mean and divide by the standard deviation for each dimension. Because the base distribution of the NF is a Gaussian, standardizing the data simplifies the mapping that the NF has to learn. Using the \verb|AbstractBijection| class in FlowJAX \cite{ward2026flowjax}, we are able to constrain the flow correctly, so that the support intervals of each parameter are obeyed and the hierarchical ordering of $m_1\geq m_2$ is adhered to.

We make use of GPU optimization by implementing all calculations in JAX \cite{Bradbury2018jax}, where the specific NF functionality is provided by FlowJAX \cite{ward2026flowjax}. Through manual testing\footnote{E.g. when testing a larger model with 14 knots and layers, the loss evolution is noisy and spiky, suggesting overfitting due to the large network complexity; other base distributions do not yield any better results and other flow types such as a Masked Autoregressive Flow \cite{Kingma2016MAF, Papamakarios2017MAF} struggle to replicate some sharp features in the parameter distributions.}, we found that a \verb|triangular_spline_flow| \cite{ward2026flowjax} with a normal distribution as the base distribution and a moderate number of 10 knots and 10 flow layers performs best. We ran training for up to 5000 epochs with a large patience of 500 using the validation loss as the stopping criterion, which usually meant training finished at around 2000 epochs. The loss function itself is the negative log-likelihood (see eq. \ref{eq:NLL}). Training on a fairly modern GPU (e.g. NVIDIA A40, A100, H100) takes less than one hour, using a batch size of 2048 or 4096. This allows for very quick retraining in the future for O4c or other, more specific applications. One such specific application could be the use of a different FAR threshold.

\section{Metrics \label{sec:evalmetrics}}

Next, we describe the metrics used to assess the NF fit and the full emulator. The results can be found in section \ref{sec:results}. Our evaluation is based primarily on two-sample tests, both for the generative NF itself and for the end-to-end emulator, similar to \cite{Callister2024Emulator}. Each two-sample test computes a statistic that quantifies the discrepancy between the two sets of samples. The two sets of samples are the test set from the injection data and a secondary set to be tested (e.g. NF samples). Under the null hypothesis that the two sets of samples are drawn from the same underlying distribution, the test statistic follows some distribution rather than taking a single value. The p-value is then the probability, under this null distribution, of obtaining a statistic at least as large as the observed one. We reject the null hypothesis at a significance level of $\alpha=0.05$.

The interpretation of p-values depends on the sample size. The rejection probability of the null hypothesis approaches 1 asymptotically as the number of samples increases \cite{Vaart1998,Lehmann2022}. On the contrary, too few samples will lead to acceptance of relatively large discrepancies, because the test is unable to capture the discrepancy at that set size. We thus perform a prospective power analysis to set a \textit{lower} limit for the number of samples required during the evaluation \cite{Cohen2013}. The details of this estimate can be found in the Appendix \ref{appendixD:twosamplepower}. We find that $\geq 30000$ samples are necessary on average to meet the desired power; this requirement is set by PQMass (see Sec. \ref{sec:pqmass}), the most demanding of the four tests. By setting a split ratio of 0.9 between training and test data we obtain $\approx 47000$ test samples. This exceeds the number of samples required. We describe the four two-sample tests used in the following.

\subsection{Marginal Tests}

We use the Kolmogorov-Smirnov (KS) \cite{Kolmogorov1933,Smirnov1948} and Anderson-Darling (AD) \cite{Anderson1952,Pettitt1976} tests to assess marginal one-dimensional distributions. Both tests are based on the cumulative distribution function (CDF). The AD statistic weights the tails of the distributions more heavily than the KS test, which essentially looks for the largest gap between the two empirical CDFs over the complete domain. Combining these tests therefore probes the central and tail regions of the marginal distributions. For the KS test, the null distribution required to compute the p-value can be obtained from a closed-form approximation. The AD test null is estimated using $N = 1000$ permutations. For both tests, we use the SciPy implementation \cite{Virtanen2020scipy}. The KS and AD tests cannot probe the full-dimensional parameter space at once, but only marginal distributions. We therefore use additional two-sample tests.

\subsection{PQMass Test}\label{sec:pqmass}

The PQMass test \cite{Lemos2024PQM} compares distributions in the full parameter space with a $\chi^2$-test using Voronoi cells. It tessellates the high-dimensional space into small non-overlapping subregions. In each region, the actual counts of each dataset are recorded. Under the null hypothesis that both datasets originate from the same underlying distribution, the expected counts in each region are derived within each cell. A $\chi^2$ test of homogeneity then evaluates the alignment of the count distributions. We use a custom JAX implementation \cite{Gassert2026PQMjax}.

\subsection{Classifier Test}

Following \cite{Lopez2016C2ST}, we also train a binary classifier (a fully connected network) on the two datasets. We assign the test data label 1 and the NF samples label 0. If the classifier is able to distinguish between the two datasets after training, this may imply that the generative model is not reproducing the data well. On the other hand, if the binary classifier performs no better than random guessing, the generative model is reproducing the data well. We assess the classifier via the area under the curve (AUC) of the receiver operating characteristic (ROC). The latter describes the curve of the true positive rate (TPR) vs. the false positive rate (FPR). For a perfect classifier, this is a flat line at TPR = 1, with a resulting AUC of 1. A diagonal from $(0,0)$ to $(1,1)$ represents random guessing and leads to an AUC of 0.5. Similarly, the accuracy (ACC) reports the fraction of correctly identified labels, which we also report but do not compute a p-value for. The classifier two-sample test (C2ST) also works in arbitrary dimensions.

\subsection{Combining Two-Sample Tests}

 The individual two-sample tests described comprise a comprehensive set of tests we use to evaluate the NF and end-to-end emulator. To combine the significance of the p-values, we use the Cauchy combination method \cite{Liu2019}, which does not require independence between the p-values. Note that individual p-values below the significance level of 0.05 do not necessarily imply that the null hypothesis should be rejected. Instead, if \textit{individual} p-values are to be assessed, the significance level for each test can be set with the Holm–Bonferroni method \cite{Bonferroni1936, Holm1979pvals}. By ordering all $m$ individual p-values $P_1 < P_2 < \dots < P_m$, one finds the smallest $k$ for which $P_k > \alpha/(m+1-k)$ and rejects the null hypotheses associated with $P_1,\dots,P_{k-1}$, retaining the remainder.

\section{Results \label{sec:results}}

We present our results below. First, the NF is evaluated as a generative model, using the two-sample tests detailed above. We then test the end-to-end emulator for predicting detection probabilities and thereby estimating integrated quantities such as the detection efficiency for different population models. This includes discussing the over-predictions and the emulator's ability to reproduce all GW parameter distributions of detected events after sampling from various population models. 
In what follows, all metrics and figures shown display the results for the O4a emulator, but the results are nearly identical for the O4b case. The main difference is that, on average, the O4b emulator predicts slightly increased detection probabilities. This is in line with the expectation that the addition of the Virgo detector improves the observational capabilities of the GW network.

\subsection{Normalizing Flow Density Estimation \label{subsec:nfstandaloneeval}}

As the NF is a generative model, we can sample fake events from it and obtain the joint distribution of the 13 parameters. We compare these distributions subsequently against the distributions of the test data, which consists of the detected injection data only. As a basic check, we ensure that the summary statistics (mean, standard deviation and quantiles of each distribution) and the correlations between different GW parameters are similar. We find good overall agreement, most importantly, for the summary statistics and correlations of the masses and redshifts, the parameters of highest importance for determining detection. Figures showing the agreement between the summary statistics from the test set and the NF samples can be found in the Appendix \ref{subsec:absRelDifferenceSSTandCorr}.

\begin{figure}[t]
    \centering
    \includegraphics[width=0.45\textwidth]{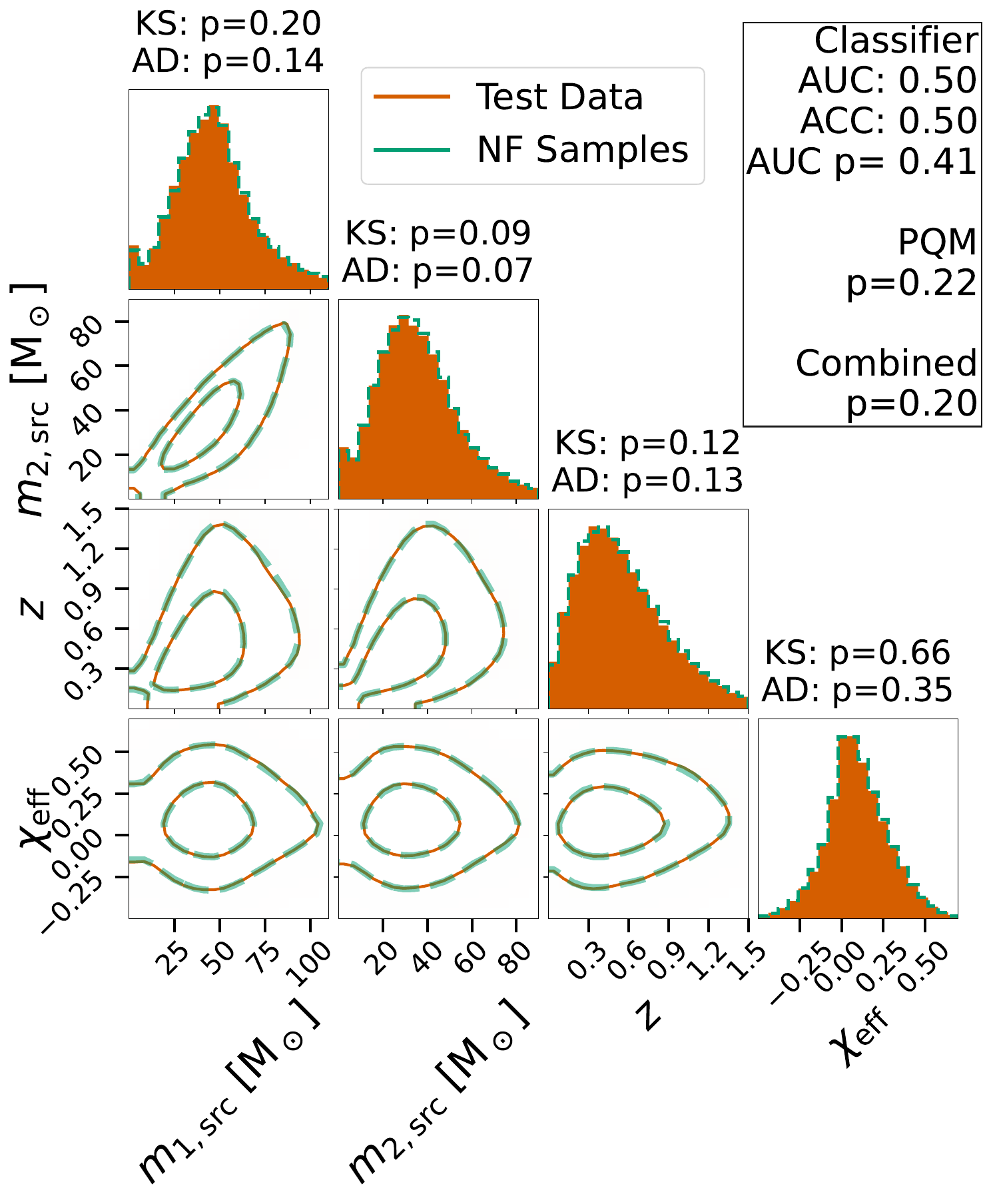}
    \caption{Joint and marginal distributions for a subset of the fitted parameters (50\% and 90\% credible intervals). Green displays the data sampled from the NF, whereas orange depicts the ground-truth test data. Only a subset of the 13 parameters is displayed, including masses, redshift and effective spin. We show the p-values for all two-sample tests and the AUC and ACC classifier metrics, indicating that it is not possible to distinguish between the NF samples and test data.\label{fig:InjectionsVSNFSamples4D}}
\end{figure}

As a next evaluation step, we consider joint and marginal distributions of the 13 parameters, overlaying the NF sample distributions with the test-set distributions. We show the full 13-dimensional plot in the Appendix \ref{subsec:jointAndMarginal13d} and focus here on masses, redshifts, and (projected) spins in Figure \ref{fig:InjectionsVSNFSamples4D}, where masses are in the source (``src'') frame. Although we have used individual spin magnitudes during training, we combine them here using the effective spin parameter $\chi_\mathrm{eff}$, which also includes the tilt angles for a more compact representation. The effective spin parameter $\chi_\mathrm{eff}$ is defined as $\chi_\mathrm{eff} = (m_1a_1\cos\tau_1 + m_2a_2\cos\tau_2)/(m_1+m_2)$.

Focusing on marginal distributions only, we see that the flow-sampled distributions match well for all parameters. The KS and AD tests for the parameters suggest that one cannot reject the hypothesis that the distributions match given their associated p-values. This is also confirmed by the p-values for the high-dimensional two-sample tests and the combined p-value over all tests. These findings establish that the NF fits the data well and captures the individual and joint distributions.


\subsection{Over-predictions \label{subsec:plausibilitychecks}}

\begin{figure}[!ht]
    \centering
    \includegraphics[width=0.48\textwidth]{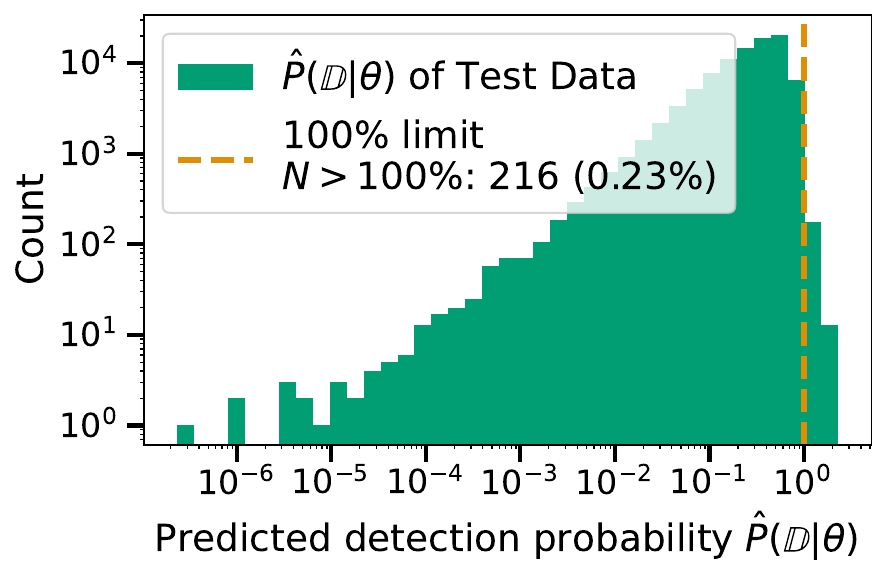}
    \caption{Distribution of over-predictions for the test set. A tiny fraction ($< 0.25\%$) of predictions overshoot to values $P(\mathbb{D} | \theta) > 1$. The maximum predicted detection probability is $P(\mathbb{D}| \theta) \approx 2.3$, while the majority of over-predictions exceed the limit  by a smaller amount.}
    \label{fig:Overpredictions}
\end{figure}

The Bayes' ratio approach for evaluating the detection probability is impacted by a finite amount of training data and finite model capacity, which may lead to over-predictions (predicted detection probabilities exceeding the limit of 1). We show the distribution of detection probabilities from test data (which only includes detections) in Figure \ref{fig:Overpredictions}. The vast majority of predicted detection probabilities lie within the 0-1 range, while $<0.25\%$ exceed 1.

\begin{figure}[t]
    \centering
    \includegraphics[width=0.48\textwidth]{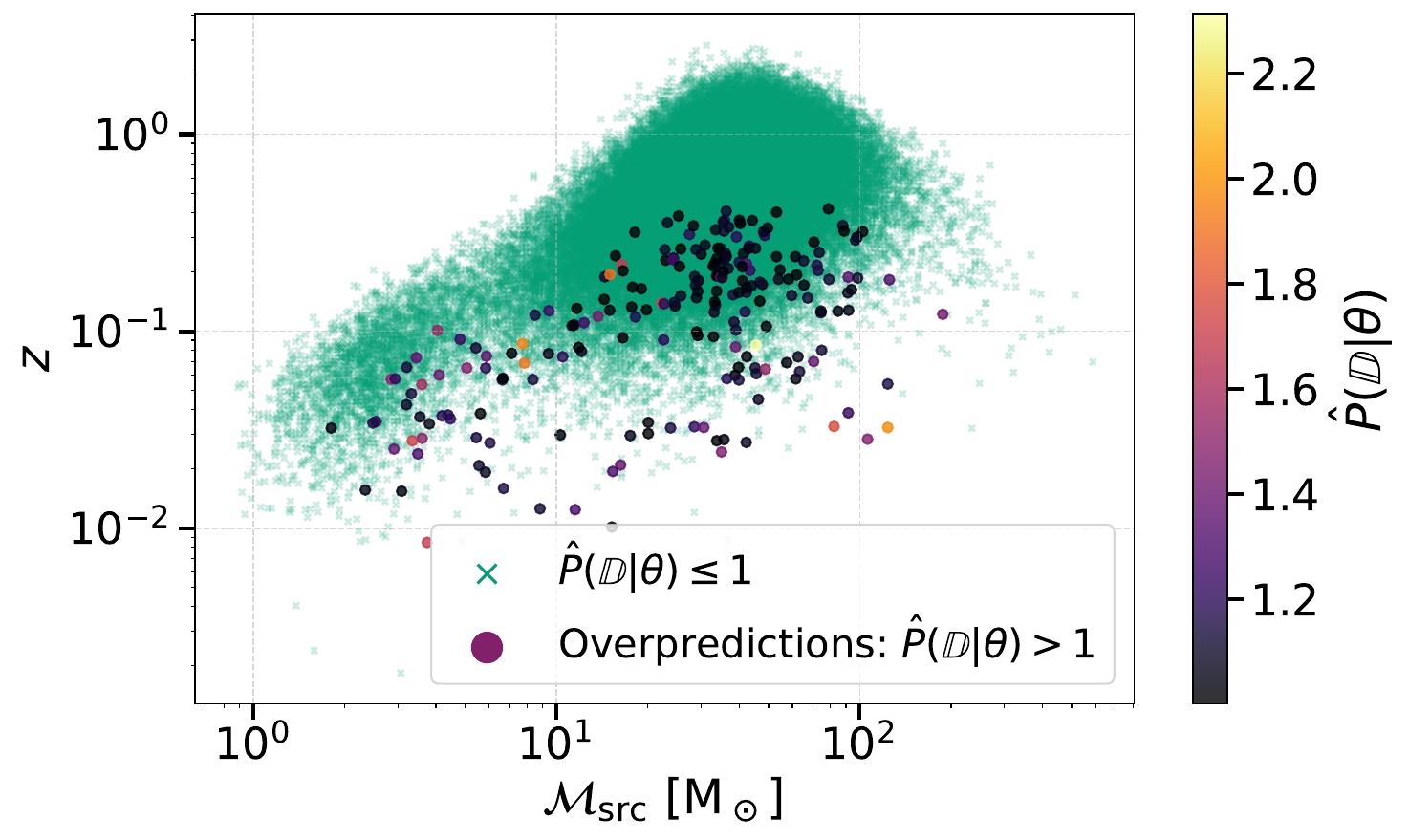}
    \caption{Predicted detection probabilities for varying chirp mass and redshift: Green crosses indicate valid detection probabilities below 100\%, while the color-coded circles depict the over-predictions. Over-predictions happen in a parameter region where high detection probability is reasonable (low redshifts, $z < 0.5$, and/or large masses).}
    \label{fig:PdetWParams}
\end{figure}

A natural question that arises is where the over-predictions lie in the parameter space. In particular, we show the dependence of the predicted detection probabilities on the chirp mass $\mathcal{M}_\mathrm{src}$ and the redshift $z$ in Figure \ref{fig:PdetWParams}. The GW strain amplitude $h$ scales as $h \propto \frac{\mathcal{M}_\mathrm{det}^{5/3}}{d_L}f^{2/3}$ \cite{Sathyaprakash2009}, where $d_L$ denotes the luminosity distance, $f$ the frequency and $\mathcal{M}_\mathrm{det}$ refers to the chirp mass in the detector frame. The latter is defined as $\mathcal{M} = \frac{(m_1m_2)^{3/5}}{(m_1 + m_2)^{1/5}}$ and the conversion from the source frame to the detector frame is carried out by multiplying by $1+z$. Detectability is therefore expected to increase at lower redshifts and/or higher masses. Finding over-predictions in such a space therefore does not indicate any major problem with the method and implementation: while the over-predictions are invalid, they appear in parts of the parameter space where the LVK detector network would detect these events with nearly 100\% probability.  We find no other significant dependency between the over-predicted detection probability and the binary merger parameters. Combining the facts that the frequency of these over-predictions is relatively low and in most cases only a minor overshoot above 100\% occurs (81\% and 94\% of over-predictions are below a predicted detection probability of 1.25 and 1.5, respectively), we conclude that the over-predictions do not pose a significant problem. A possible origin of these over-predictions is the following chain: fluctuations in the estimation of the density, stemming from the numerical approximation with a finite amount of training data and model capacity, which are subsequently inflated by the Bayes' ratio.

Distributions of predicted detection probabilities for all events from the injection data set (not only the detected events as in the test set) and the corresponding over-predictions show trends similar to those in the figures above, albeit with the clear difference that a larger fraction has a lower predicted detection probability and even fewer over-predictions (less than 0.05\%) occur. At first, one might expect that these non-detections (in the injection campaign) should all receive a very low predicted detection probability. However, detection is decided based on a FAR threshold that contains a stochastic noise contribution from detector noise. For events with very large (very low) SNR, detection (non-detection) is essentially certain. Near the detection threshold, however, the same binary parameters can yield a detection or non-detection depending on the noise.  Since the injection campaign is designed to probe this transition region, it is no surprise that our proposed method attributes moderate to large detection probabilities to a significant number of samples in the non-detected injection set. As a concrete example, of the events that were not detected in the injection campaign, about 27\% have a predicted detection probability of more than 25\%. When comparing this with the O3 emulator \cite{Callister2024Emulator}, the distributions and predicted detection probabilities exhibit nearly identical behavior (except for the fact that no over-predictions can occur, which is due to the different approach in the ML technique used). 

With the arguments presented here, we conclude that due to the low frequency, statistical fluctuation as possible origin, and affected parameter space of the over-predictions, their occurrences likely do not pose significant problems for our method. Therefore, we implement a simple warning in our code, displaying a message about a suspiciously large fraction (more than 1\%) of over-predictions in case this happens. Our code defaults to clipping over-predictions to a predicted detection probability of 1.

\subsection{Detected Populations and Detection Efficiencies\label{subsec:nffoundeval}}

We have established that the NF fit is satisfactory in section \ref{subsec:nfstandaloneeval} and investigated over-predictions. Although these evaluations indicate that the method works, we provide additional tests relevant for population studies. We {test the end-to-end emulator in the following}. We draw the 13 binary merger parameters $\theta_e$ per event $e$ for many events from the same population model $\Lambda_\mathrm{inj}$ as used in the injection campaign, evaluate the associated predicted detection probability $P(\mathbb{D} | \theta_e)$ for each event {using the NF emulator} and probabilistically accept events as detected \footnote{Probabilistically accepting events means that we draw a random number $r \in [0,1]$ for every predicted detection probability $P(\mathbb{D} | \theta_e)$ associated with an event and mark it as detected if the predicted detection probability is larger than the random number, i.e. if $P(\mathbb{D} | \theta_e) > r$.} (see \ref{subsec:appendixB_DrawingSamples} for details on how we draw the samples from the probability density functions). This is repeated until a sufficient number of detections is recorded (the same number of detections as the test set size). These detected events can then be used to apply similar tests as described in section \ref{sec:evalmetrics}, to confirm a good reproduction of the distributions of parameters for detected events from the injection set, evaluation of over-predictions and parameter dependencies for detectability, and a correct prediction of the detection efficiency. We refer to the detected events from the procedure just described as ``emulator detections'', in order to distinguish them from the detected events in the injection campaign.

\begin{figure}[t]
    \centering
    \includegraphics[width=0.48\textwidth]{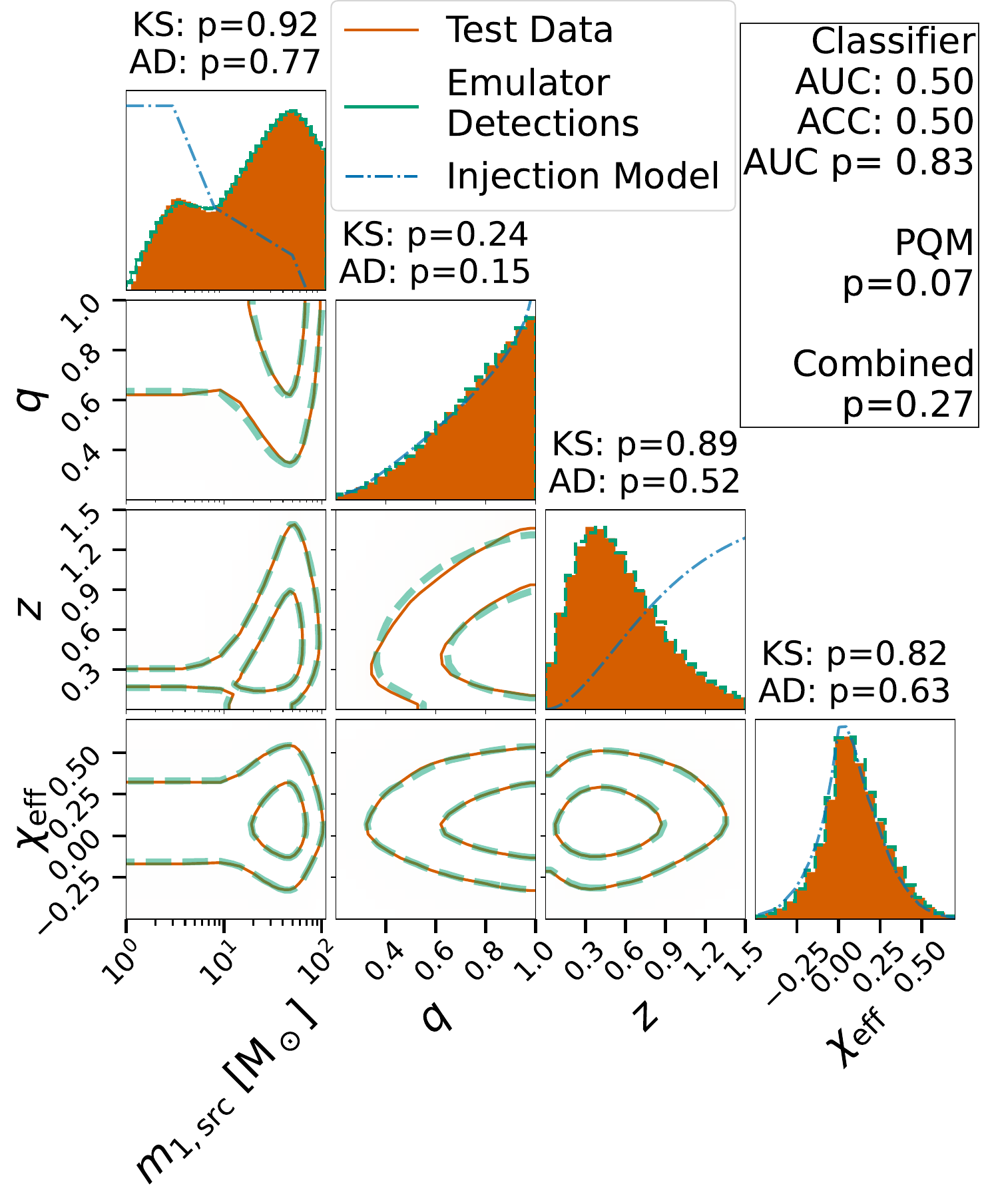}
    \caption{Joint and marginal parameter distributions (50\% and 90\% credible intervals) for four main binary merger parameters (primary mass, mass ratio $q$, redshift and effective spin). Orange represents the test set (ground truth), while green shows the emulator detections, generated by probabilistically accepting sampled events as detected using the end-to-end emulator. The injection model probability density functions are shown with blue dash-dotted lines. We show the classifier metrics and PQMass test p-value, indicating good agreement between the two distributions for all parameters.}
    \label{fig:InjectionsVSFoundSamples4D}
\end{figure}

We begin with the parameter distributions for the emulator detections. A quantitative comparison of the summary statistics shows good agreement between the distributions. In Figure \ref{fig:InjectionsVSFoundSamples4D} we show the joint and marginal distributions for a 4D subset of parameters. Since we initially draw parameters from the population model used in the injection campaign, we also show the injection model probability density functions for the individual parameters. In addition, we show the mass ratio $q = m_2/m_1$ instead of the secondary mass $m_2$.

The marginal and joint distribution tests suggest that the null hypothesis cannot be rejected and the classifier is again unable to meaningfully distinguish between the datasets. A visual inspection of the joint and marginal parameter distributions confirms the quantitative analysis, and, in combination, we conclude once again that the distributions are recovered correctly. In this case as well, we find that the number of over-predictions is low ($< 0.25$\%) and the over-predictions occur in plausible regions of the parameter space. With the arguments presented above, we conclude that this is not a significant issue.

As one of the main applications of this emulator is hierarchical Bayesian inference for compact object population analysis, we perform one final test in this context. With the emulator, we can estimate the detection efficiency used in compact object population analysis. We first confirm that the emulator estimate of the detection efficiency for the injection model matches the MC estimate and then compare MC-IS against the emulator for a variety of differently parameterized population models. The detection efficiency calculated from the injection data via MC integration yields a value of $\hat{\xi}^\mathrm{MC}(\Lambda_\mathrm{inj}) = \hat{P}(\mathbb{D}) = (1.125 \pm 0.002) \times 10^{-3}$. The estimated detection efficiency from the emulator detections is $\hat{\xi}^\mathrm{NF}(\Lambda_\mathrm{inj}) = \hat{P}(\mathbb{D}) = (1.122\pm 0.004) \times 10^{-3}$. These results are consistent with each other within $1\sigma$ and provide further support for our method and implementation.

Finally, we repeat the above procedure of generating emulator detections to estimate the detection efficiency, but this time altering the population model $\Lambda_i$ for each iteration. Specifically, in each iteration, we first draw a new set of population-model parameters from the standard priors given in Appendix \ref{subsec:appendixB_GWTC5Model}. We then use these parameters to draw the source population for that iteration, including the component masses, spins, and redshifts. Using a standard GWTC-5-like population model \cite{GWTC5Population2026}, we adopt a \textsc{Broken Power Law + 2 Peaks} model for the primary mass and a power law conditioned on the primary mass for the secondary masses. The spin magnitudes follow identically truncated Gaussians and are assumed to be independent. Spin tilts are modeled as a mixture of truncated Gaussians and an isotropic distribution. Redshifts are distributed proportionally to the comoving merger rate density following a power-law rate evolution. Details of the population model can be found in Appendix \ref{subsec:appendixB_GWTC5Model}.

\begin{figure}[t]
    \centering
    \includegraphics[width=0.48\textwidth]{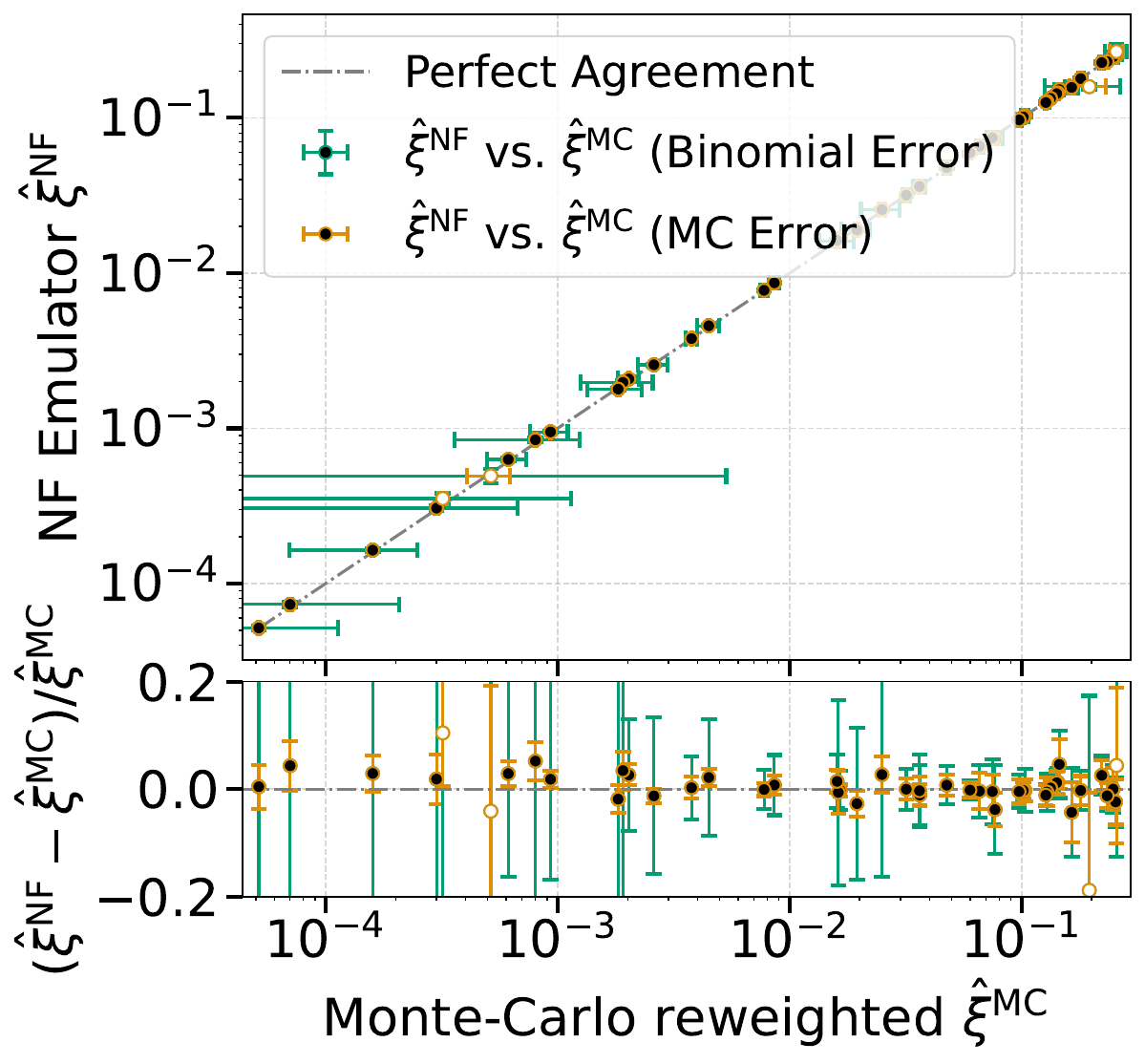}
    \caption{\emph{Upper panel:} Estimated detection efficiencies via Monte Carlo importance sampling of the injection set ($\hat{\xi}^\mathrm{MC}$) versus our proposed NF method ($\hat{\xi}^\mathrm{NF}$). The NF emulator detection efficiencies are averages over an ensemble of eight fully trained NFs, each individually predicting the detection efficiency for 70,000 detected samples. The median relative deviation between the two estimated detection efficiencies is $\approx 1.2\%$. White circles indicate population models where $N_\mathrm{eff}<4N_\mathrm{obs}$. 
    \emph{Lower panel:} residuals between the two estimates as a function of $\hat{\xi}^\mathrm{MC}$. In all cases the estimated detection efficiencies lie within the $1\sigma$ errors. For regions of low $N_\mathrm{eff}$ both the binomial and MC error dominate the combined error. Elsewhere, combined errors of NF and MC are similarly large, while the binomial error is still larger than the NF error. }
    \label{fig:DetectionEfficiencies}
\end{figure}

To incorporate some notion of emulator uncertainty, we train an ensemble of emulators (consisting of 8 NFs) with varying network capacities and initial seeds. We then compare the MC-IS detection efficiency obtained from the injection data with the averaged estimate from the emulators. The MC-IS estimate of the detection efficiency has an associated error as described in section \ref{subsec:deteff}. The result is summarized in Figure \ref{fig:DetectionEfficiencies}. We find good agreement between the two estimates of the detection efficiency within their combined uncertainty. The combined uncertainty consists of the emulator ensemble standard deviation and the MC-IS standard deviation added in quadrature. The MC-IS estimate dominates the combined error in all cases when using the binomial error estimate. Particularly for lower detection efficiencies, this also holds for the standard MC-IS error estimate, while for high $N_\mathrm{eff}$ the errors between the NF ensemble and MC-IS are of the same order. The increased variance in MC-IS is due to the number of effective samples $N_\mathrm{eff}$ used in the calculation of the error, which may become small and increase it significantly, particularly for peaked population models.

We note that for the number of effective samples $N_\mathrm{eff}$ needed to obtain an unbiased estimate of the detection efficiency, different thresholds have been proposed in the literature, ranging from $N_\mathrm{eff} \gg  4 N_\mathrm{events}$ \cite{Farr2019,Essick2022Neff} to a steeper scaling of $N_\mathrm{eff} \gg  N_\mathrm{events}^\alpha$ with $1.5 \leq \alpha \leq 2$ \cite{Talbot2023}. For a significant number of population models, not even the lowest threshold is met, indicating that some of these population models may not allow an accurate estimate of the detection efficiency via the MC-IS method. Consequently, for these population models, extremely large uncertainties are observed, as is clear in Figure \ref{fig:DetectionEfficiencies}.

\begin{figure}[t]
    \centering
    \includegraphics[width=0.48\textwidth]{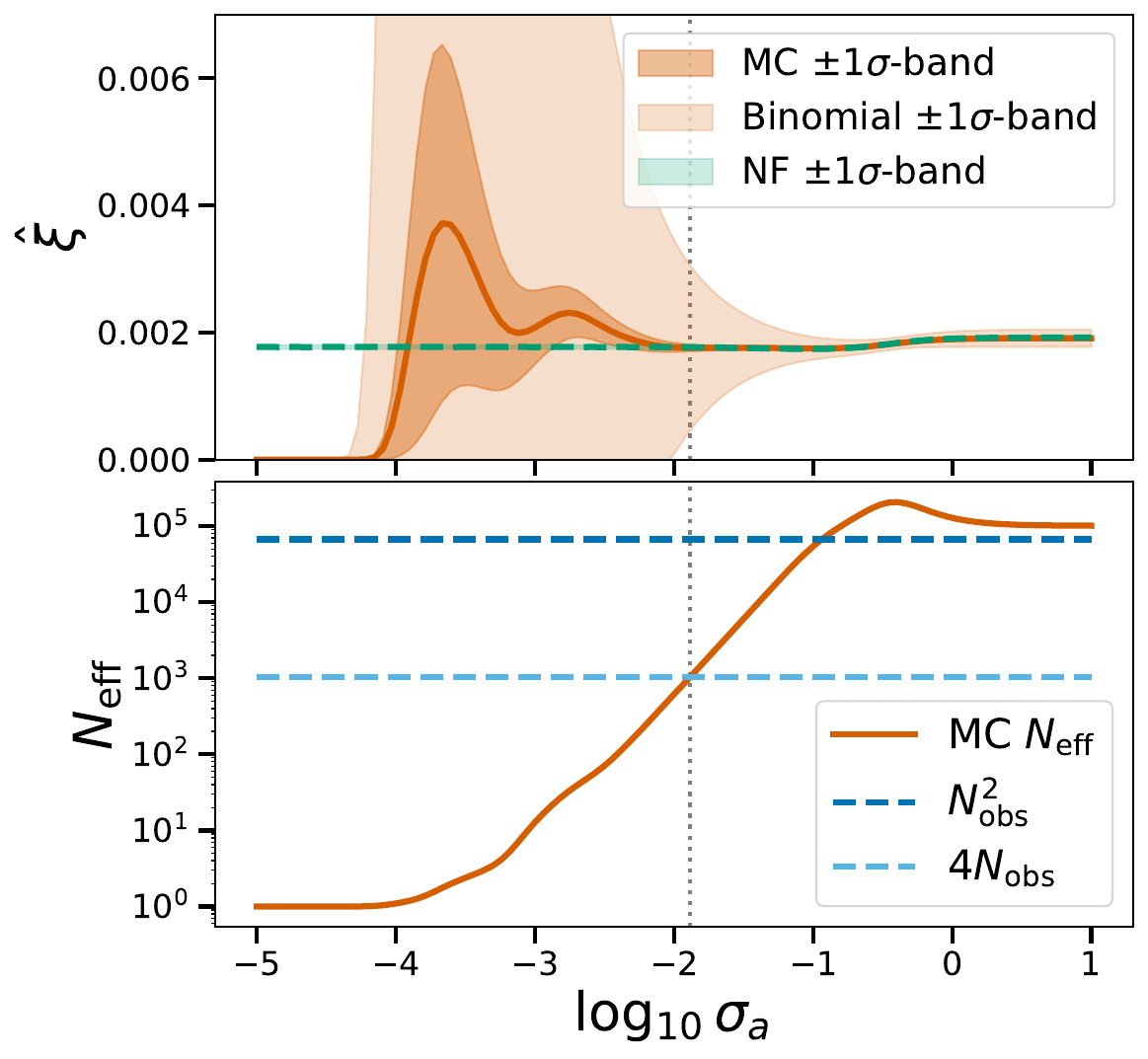}
    \caption{\emph{Upper panel:} Estimated detection efficiencies via Monte Carlo importance sampling of the injection set versus our proposed NF method. We fix all population model parameters except the width of the Gaussian spin magnitude $\sigma_a$ (see Table \ref{tab:bpl2p_default}). The NF emulator detection efficiencies are averages over an ensemble of eight fully trained NFs, each individually predicting the detection efficiency for 70,000 detected samples. Shaded bands indicate $\pm1\sigma$ error bands. The light orange indicates the binomial estimate of the error as in eq. \ref{eq:xi_Neff_err}, while the darker orange shade shows the MC-IS standard deviation. \emph{Lower panel:} Evolution of $N_\mathrm{eff}$ for the same population models as in the upper panel.}
    \label{fig:DetectionEfficienciesSigmaChi}
\end{figure}

We also show how specific population models can lead to a collapse of the MC-IS estimate of the detection efficiency in Figure \ref{fig:DetectionEfficienciesSigmaChi} and compare against the NF emulator. Here, we fix all population model parameters at a fiducial value, except for the width of the spin magnitude distribution, $\sigma_a$, as in \cite{Callister2024Emulator}. See Appendix \ref{subsec:appendixB_GWTC5Model} for details on the population model and fiducial population parameters. With an increasingly tighter spin magnitude distribution, the detectability of events should not change drastically. Therefore, the inferred detection efficiency should stay approximately constant. As the width decreases, the estimated detection efficiency using the NF emulator stays approximately constant as expected. In contrast, the MC-IS estimate strongly varies below $\sigma_a \approx 10^{-2}$. This is exactly the regime where $N_\mathrm{eff}$ drops below $4N_\mathrm{obs}$. We show the evolution of $N_\mathrm{eff}$ in the lower panel.

Figures \ref{fig:DetectionEfficiencies} and \ref{fig:DetectionEfficienciesSigmaChi} show some failure modes of the MC-IS estimates of the detection efficiency. Both the accuracy and precision of the MC-IS estimate of the detection efficiency may suffer from poorly importance-sampled estimates. The estimated detection efficiency spikes and drops to 0 for narrow spin magnitude models (Figure \ref{fig:DetectionEfficienciesSigmaChi}). For various randomly parameterized population models, the MC-IS uncertainty largely outweighs the NF emulator uncertainty (Figure \ref{fig:DetectionEfficiencies}). Therefore, the emulator presented here likely provides a more precise and flexible method to estimate the detection efficiency for a variety of population models. For extreme population models, e.g. with very small widths of Gaussian peaks, it can also be more accurate than MC-IS. All that is necessary is to specify a population model from which the parameters $\theta$ of interest can be drawn and evaluate the detection probabilities. The detection efficiency is then, as described in section \ref{subsec:deteff}, simply the mean of all predicted $P(\mathbb{D}| \theta)$.

\section{Discussion and Conclusion\label{sec:conclusion}} 

We have presented a method and the subsequent implementation of an emulator to predict the detection probability $P(\mathbb{D} | \theta)$ of binary compact object mergers given their parameters $\theta$ for a LIGO/Virgo/KAGRA O4 detector network configuration using the injection campaign as training and test data. We make both the training code and the fully trained networks publicly available and provide them as one installable package \footnote{\href{https://github.com/jgassert/graviflows}{``graviflows'' package on github}}. A simple guide on how to use the code is provided through our \href{https://jgassert.github.io/graviflows/}{``graviflows'' documentation}. We evaluated our emulator using various metrics and, most importantly, applied it to different binary merger population models to ensure an accurate and precise estimate of the detection efficiency, as population inference is a standard application for such an emulator. The two-sample tests cannot distinguish between the injection-data and emulator-recovered parameter distributions, neither when samples are drawn directly from the NF nor when they are generated by applying the emulator to artificially sampled binary parameters.

One of the main motivations for developing a detection probability estimator is to stabilize hierarchical Bayesian population inference, which may suffer dramatically from MC-IS variance when estimating the detection efficiency. This is a required term in the hierarchical Bayesian inference formalism. Redoing or extending the injection campaign to obtain more samples and therefore increase the number of effective samples is not viable, as this is labor-intensive and requires significant compute time \cite{Essick2025Injections}. Some other approaches to estimate detection efficiencies have been proposed and used in the literature \cite{Wong2021CE,Mould2022GWPops,ChapmanBird2023}. Earlier analytical predictions of the selection function via the SNR of merger events \cite{Essick2023SemiEstimates,Gerosa2024Recipes} have been shown to not reliably reproduce all parameter distributions by \cite{Callister2024Emulator}. Most recently, however, \cite{Vijaykumar2026FFSemianalytic} show that they can improve upon these results and infer the selection function more faithfully. There exists at least one example of a physically motivated selection function, presented by \cite{LorenzoMedina2025Selection} for O3, which, however, only works on a subspace of parameters.

Other works have used similar ML-based approaches to estimate the detection probability \cite{Gerosa2020Selection, Talbot2022, Callister2024Emulator}. 
We extend these approaches by treating the problem as a density estimation process trained on the injection data, using a more flexible, expressive, and generative model, updating to the latest observing run O4 and simplifying the usage of the fully trained model.

For our method, the evaluation of 1 million binary merger event parameters on a personal laptop (no GPU, no just-in-time code compilation) requires less than three seconds, we believe that computing resources are not a limiting factor in any scenario. Evaluating all $\approx 433$ million injections generated during the O4a injection campaign would take about 20 minutes. {Using a GPU and code compilation can significantly decrease run time.}

Although our emulator enables fast ($\mathcal{O}(0.01s)$ for 1 million samples after compilation) and accurate estimation of detection probabilities for the latest observing runs, it is not perfect. We observe a small fraction ($\mathcal{O}(0.1\%)$) of over-predictions in a parameter space where detection is virtually guaranteed, rendering this a negligible side effect of this method. The emulator interpolates smoothly between the support points given by the training data. Under the assumption that detectability varies smoothly over the parameter space -- which is physically motivated by the smooth dependence of detector sensitivity on binary parameters -- the interpolation should be well-behaved in most cases.

In our method, we still discard potentially relevant information:  the FAR. Future improvements could use this information to train a conditional NF instead, conditioned on the FAR. This procedure would not only encode more information through the FAR but also allow all injections to be used instead of discarding the majority of them (all non-detections). This could likely further improve the detection probability estimate. We plan to update the package for the O4c observing run once the corresponding injection campaign is published.

\begin{acknowledgments}

We thank Katie Breivik for guidance about publishing the code and Reed Essick and Daniel Gruen for useful discussions. AP is supported by NSF Grant No. 2308193 and National Aeronautics and Space Administration under Grant No. 22-LPS22-0025.

This research has made use of data or software obtained from the Gravitational Wave Open Science Center (gwosc.org), a service of the LIGO Scientific Collaboration, the Virgo Collaboration, and KAGRA. This material is based upon work supported by NSF's LIGO Laboratory which is a major facility fully funded by the National Science Foundation, as well as the Science and Technology Facilities Council (STFC) of the United Kingdom, the Max-Planck-Society (MPS), and the State of Niedersachsen/Germany for support of the construction of Advanced LIGO and construction and operation of the GEO600 detector. Additional support for Advanced LIGO was provided by the Australian Research Council. Virgo is funded, through the European Gravitational Observatory (EGO), by the French Centre National de Recherche Scientifique (CNRS), the Italian Istituto Nazionale di Fisica Nucleare (INFN) and the Dutch Nikhef, with contributions by institutions from Belgium, Germany, Greece, Hungary, Ireland, Japan, Monaco, Poland, Portugal, Spain. KAGRA is supported by Ministry of Education, Culture, Sports, Science and Technology (MEXT), Japan Society for the Promotion of Science (JSPS) in Japan; National Research Foundation (NRF) and Ministry of Science and ICT (MSIT) in Korea; Academia Sinica (AS) and National Science and Technology Council (NSTC) in Taiwan.

Large Language Models (Claude Sonnet, Opus 4.8 and Opus 5) have been used in the preparation of this work, mostly to assist in code writing and review. All generated code and outputs were subsequently verified, tested, and validated by the authors.

The authors acknowledge the IT Support Team of the Faculty of Physics at LMU for maintaining and supporting the HPC, storage, and computational infrastructure used in this work. This research used resources provided by the Vera Cluster at the Pittsburgh Supercomputing Center (PSC).

\end{acknowledgments}

\appendix

\section{Models and Sampling\label{sec:appendixA_samplpdfs}}

\subsection{Probability Density Functions of the Injection Model\label{subsec:appendixB_pdfs}} 

We list the 13 binary merger parameters, their symbols and probability density functions of the injection model in Table \ref{tab:injection_priors}. The original population model definition is stated in \cite{Essick2025Injections}.

\begin{table*}[t]
    \caption{\label{tab:injection_priors}Parameters $\theta$ of the injection campaign
    and their probability density functions.}
    \begin{ruledtabular}
        \begin{tabular}{lll}
        Parameter & Symbol & Probability density function \\ \hline
        Primary source-frame mass   & $m_{1,\mathrm{src}}$ & eq~(\ref{eq:m1prior}) \\
        Secondary source-frame mass & $m_{2,\mathrm{src}}$ & $2 m_{2,\mathrm{src}}/(m_{1,\mathrm{src}}^{2}-1)\,\Theta(1 \leq m_{2,\mathrm{src}} \leq m_{1,\mathrm{src}})$ \\
        Spin magnitudes             & $a_{1},\,a_{2}$ & $\propto e^{-2a_{i}^{2}}$ \\
        Spin tilts                  & $\cos\tau_{1},\,\cos\tau_{2}$ & $0.3\,(1+\cos\tau_{i})^{3}/4 + 0.7/2$ \\
        Azimuthal spin angles       & $\phi_{1},\,\phi_{2}$ & $\mathcal{U}(0, 2\pi)$ \\
        Inclination                 & $\cos\theta_{JN}$ & $\mathcal{U}(-1, 1)$ \\
        Polarization                & $\psi$ & $\mathcal{U}(0, \pi)$ \\
        Right ascension             & $\alpha$ & $\mathcal{U}(0, 2\pi)$ \\
        Declination                 & $\sin\delta$ & $\mathcal{U}(-1, 1)$ \\
        Redshift                    & $z$ & $\propto dV_{c}/dz$, $z \in [0,3)$ \\
        \end{tabular}
    \end{ruledtabular}
\end{table*}

The primary mass $m_1$ is distributed according to a broken five-piece power law:
\begin{widetext}
    \begin{equation} \label{eq:m1prior}
    p(m_{1,\mathrm{src}} \mid \Lambda_{\mathrm{inj}}) \propto
        \begin{cases}
        0 & m_{1,\mathrm{src}} < 1\,\mathrm{M}_\odot \\
        1 & 1\,\mathrm{M}_\odot \leq m_{1,\mathrm{src}} < 3\,\mathrm{M}_\odot \\
        \left(\dfrac{m_{1,\mathrm{src}}}{3}\right)^{-4} & 3\,\mathrm{M}_\odot \leq m_{1,\mathrm{src}} < 8\,\mathrm{M}_\odot \\
        \left(\dfrac{8}{3}\right)^{-4}\left(\dfrac{m_{1,\mathrm{src}}}{8}\right)^{-1} & 8\,\mathrm{M}_\odot \leq m_{1,\mathrm{src}} < 50\,\mathrm{M}_\odot \\
        \left(\dfrac{8}{3}\right)^{-4}\left(\dfrac{50}{8}\right)^{-1}\left(\dfrac{m_{1,\mathrm{src}}}{50}\right)^{-4} & 50\,\mathrm{M}_\odot \leq m_{1,\mathrm{src}} < 200\,\mathrm{M}_\odot \\
        \left(\dfrac{8}{3}\right)^{-4}\left(\dfrac{50}{8}\right)^{-1}\left(\dfrac{200}{50}\right)^{-4}\left(\dfrac{m_{1,\mathrm{src}}}{200}\right)^{-1} & 200\,\mathrm{M}_\odot \leq m_{1,\mathrm{src}} < 1000\,\mathrm{M}_\odot \\
        0 & m_{1,\mathrm{src}} \geq 1000\,\mathrm{M}_\odot
        \end{cases}
    \end{equation}
\end{widetext}

No tidal effects are included in the injection campaign, and orbits are assumed to be quasi-circular. 

\subsection{GWTC-5-like Population Model\label{subsec:appendixB_GWTC5Model}}

Throughout this work, we have used a GWTC-5-like \textsc{Default BBH} population model \cite{GWTC5Population2026}. The primary-mass distribution is modeled as a \textsc{Broken Power Law + 2 Peaks} mixture of a broken power law (BP) and two left-truncated Gaussian peaks, denoted by $\mathcal{N}_\mathrm{lt}(\cdot)$, with a low-mass tapering function $S(\cdot)$ applied to the full distribution. The full mixture is given as

\begin{align}
p(m_1) &\propto \Big[
    \lambda_0 p_{\mathrm{BP}}\big(
        m_1 \mid \alpha_1, \alpha_2,
        m_{\mathrm{break}}, m_{1,\mathrm{low}}, m_{\mathrm{high}}
    \big)  \nonumber \\
 &+ \lambda_1 \mathcal{N}_{\mathrm{lt}}\big(
        m_1 \mid \mu_1, \sigma_1,
        \mathrm{low}=m_{1,\mathrm{low}}
    \big) \nonumber \\ 
& + (1-\lambda_0-\lambda_1)
    \mathcal{N}_{\mathrm{lt}}\big(
        m_1 \mid \mu_2, \sigma_2,
        \mathrm{low}=m_{1,\mathrm{low}}
    \big)
\Big] \nonumber \\ 
    & S\big(
        m_1 \mid \mathrm{low}=m_{1,\mathrm{low}},
        \delta_{m,1}
    \big),
\end{align}

where the broken power law is defined as
\begin{widetext}
    \begin{equation}
        p_\mathrm{BP}(m_1|\alpha_1, \alpha_2, m_\mathrm{break}, m_{1,\mathrm{low}}, m_\mathrm{high}) = \frac{1}{N}
        \begin{cases}
            \left(\dfrac{m_1}{m_\mathrm{break}}\right)^{-\alpha_1} & m_{1,\mathrm{low}} \leq m_1 \leq m_\mathrm{break}, \\[6pt]
            \left(\dfrac{m_1}{m_\mathrm{break}}\right)^{-\alpha_2} & m_\mathrm{break} \leq m_1 \leq m_\mathrm{high} \ .
        \end{cases}
    \end{equation}
    The normalization of the broken power law is computed as
    \begin{equation}
        N = m_\mathrm{break}\left[\frac{1-(m_{1,\mathrm{low}}/m_\mathrm{break})^{1-\alpha_1}}{1-\alpha_1} + \frac{(m_\mathrm{high}/m_\mathrm{break})^{1-\alpha_2}-1}{1-\alpha_2}\right],
    \end{equation}
    and the low-mass tapering follows
    \begin{equation}
        S(m|m_\mathrm{low}, \delta_m) =
        \begin{cases}
            0 & m < m_\mathrm{low}, \\
            \left[1+\exp\left(\dfrac{\delta_m}{m-m_\mathrm{low}}+\dfrac{\delta_m}{m-m_\mathrm{low}-\delta_m}\right)\right]^{-1} & m_\mathrm{low} \leq m < m_\mathrm{low} + \delta_m, \\
            1 & m_\mathrm{low} + \delta_m \leq m .
        \end{cases}
    \end{equation}
\end{widetext}
The secondary masses are modeled with a power law, conditioned on the primary mass and independently tapered at the low-mass end

\begin{align}
    p(m_2 | m_1) \propto m_2^{\beta_q} S(m_2 | m_{2, \mathrm{low}},\delta_{m_2}) \,
\end{align}

where $m_{2,\mathrm{low}} \leq m_2 \leq m_1$. Spin magnitudes ($a_i$) follow a truncated Gaussian distribution between 0 and 1, assuming independence:

\begin{align}
    p(a_1,a_2|\mu_a, \sigma_a) = \mathcal{N}_{[0,1]}(a_1|\mu_a, \sigma_a) \mathcal{N}_{[0,1]}(a_2|\mu_a, \sigma_a)
\end{align}
The cosine spin-tilt angles are modeled using a mixture of a Gaussian distribution truncated to $[-1,1]$ and an isotropic distribution:

\begin{align}
    & p(\cos\tau_1,\cos\tau_2 | \mu_\tau, \sigma_\tau, \zeta) = \nonumber \\ & \quad \zeta \mathcal{N}_{[-1,1]}(\cos\tau_1|\mu_\tau, \sigma_\tau)\mathcal{N}_{[-1,1]}(\cos\tau_2|\mu_\tau, \sigma_\tau) + \frac{1-\zeta}{4}
\end{align}
We model the redshift evolution following the comoving merger rate density, such that

\begin{align}
    p(z|\kappa) \propto \frac{1}{1+z} \frac{\mathrm{d}V_c}{\mathrm{d} z} (1+z)^\kappa.
\end{align}
A summary of the model parameters of the \textsc{Default BBH} model, together with their priors and fiducial values, can be found in Table \ref{tab:bpl2p_default}.

\begin{table*}[t]
\caption{\label{tab:bpl2p_default}Summary of \textsc{Broken Power Law + 2 Peaks} mass model
parameters, spin model, and redshift distribution parameters, together with their priors and fiducial values.}
    \begin{tabular}{llll}
        \hline\hline
        Parameter & Description & Prior (Fig.~\ref{fig:DetectionEfficiencies}) & Fiducial (Fig.~\ref{fig:DetectionEfficienciesSigmaChi}) \\ \hline
        $\alpha_1$ & Power-law index of 1st primary-mass power law & $\mathcal{U}(-4,12)$ & 1.5 \\
        $\alpha_2$ & Power-law index of 2nd primary-mass power law & $\mathcal{U}(-4,12)$ & 5.4 \\
        $m_{\mathrm{break}}$ & Power-law break location & $\mathcal{U}(20,50)$ & $35.0\,\msun$ \\
        $\mu_1$ & Location of the first peak & $\mathcal{U}(5,20)$ & $10.0\,\msun$ \\
        $\sigma_1$ & Width of the first peak & $\mathcal{U}(0,10)$ & $1.5\,\msun$ \\
        $\mu_2$ & Location of the second peak & $\mathcal{U}(25,60)$ & $33.0\,\msun$ \\
        $\sigma_2$ & Width of the second peak & $\mathcal{U}(0,10)$ & $4.0\,\msun$ \\
        $m_{\mathrm{1,low}}$ & Lower edge of taper function & $\mathcal{U}(3,10)$ & $5.0\,\msun$ \\
        $\delta_{\mathrm{m,1}}$ & Mass range of low mass tapering & $\mathcal{U}(0,10)$ & $4.0\,\msun$ \\
        $\lambda_0,\lambda_1$ & Mixing fractions between power law and peaks & $\mathrm{Dir}(\alpha=(1,1,1))$ & 0.88, 0.06 \\
        $m_{\mathrm{high}}$ & Maximum mass for distribution (pinned to $m_{\mathrm{high}}=300\,\msun$ by default) & $\delta(m_{\mathrm{high}}-300)$ & $300\,\msun$ \\ \hline
        $\beta_q$ & Power-law index of mass-ratio power law & $\mathcal{U}(-2,7)$ & 1.0 \\
        $m_{\mathrm{2,low}}$ & Lower edge of taper function in $m_2$ & $\mathcal{U}(3,m_{\mathrm{1,low}})$ & $4.5\,\msun$ \\
        $\delta_{\mathrm{m,2}}$ & Mass range of low-mass tapering in $m_2$ & $\mathcal{U}(0,10)$ & $2.0\,\msun$ \\ \hline
        $\mu_a$ & Location of the spin magnitude distribution & $\mathcal{U}(0,1)$ & 0.2\\
        $\sigma_a$ & Width of the spin magnitude distribution & $\mathcal{U}(0.005,1)$ & $\log\mathcal{U}(10^{-5},10^{1})$ \\
        $\mu_\tau$ & Location of the Gaussian component of the $\cos\tau$ distribution & $\mathcal{U}(-1,1)$ & 0.15 \\
        $\sigma_\tau$ & Width of the Gaussian component of the $\cos\tau$ distribution & $\mathcal{U}(0.01,4)$ & 0.8 \\
        $\zeta$ & Fraction in the Gaussian component of the $\cos\tau$ distribution & $\mathcal{U}(0,1)$ & 0.7 \\ \hline
        $\kappa$ & Power-law index on comoving merger rate evolution & $\mathcal{U}(-10,10)$ & 2.5 \\
        \hline\hline
    \end{tabular}
\end{table*}

\subsection{Drawing Samples\label{subsec:appendixB_DrawingSamples}} 

Binary merger parameters are drawn from some population model $\Lambda_i$ by inverse transform sampling. $x_j$ denotes the $j$-th binary merger parameter (there are 13 binary merger parameters, as mentioned in Appendix \ref{subsec:appendixB_pdfs}) and $F_{\Lambda_i,x_j}(x) = \int_{-\infty}^{x} p(x' \mid \Lambda_i)\,\mathrm{d}x'$ denotes its CDF. $F^{-1}_{\Lambda_i, x_j}$ thus is the quantile function (inverse of the CDF). To obtain a set of $N$ samples distributed according to $p(x_j|\Lambda_i)$, we draw $N$ independent variates $\boldsymbol{c}_{x_j} = (c^{(1)}, \dots, c^{(N)})$, where each $c^{(k)} \sim \mathcal{U}(0,1)$, and apply the quantile function. This can be summarized as:

\begin{align}
    x_j^{(k)} = F^{-1}_{\Lambda_i,x_j}\!\left(c^{(k)}\right), \qquad c^{(k)} \sim \mathcal{U}(0,1),  \quad k = 1,\dots,N
\end{align}

When parameters are correlated, the conditional CDFs are used and the parameters are drawn sequentially.

\section{Two-Sample Test Power Analysis\label{appendixD:twosamplepower}} 

As described in section \ref{sec:evalmetrics}, the two-sample test power depends on the number of data points used in the tests. ``Two-sample'' refers to testing whether two sets of samples/datasets appear to have been drawn from the same underlying distribution \footnote{Elsewhere, ``sample'' refers to a data point.}. If enough samples are provided in both sets, any discrepancy $\Delta$ between the two sets can be resolved, including sampling noise \cite{Vaart1998}. It is therefore not necessarily useful to simply take as many samples as one can afford. In contrast, too few samples may impede the detection of a true discrepancy in the datasets. This is the more dangerous failure, as it risks accepting a bad NF. Following the framework presented in \cite{Cohen2013}, we therefore perform a power analysis to set a lower bound on the number of samples necessary during evaluation. We do so for each of the four two-sample tests (KS, AD, PQMass and the C2ST) individually.

We use the real injection data to perform this analysis, by splitting it into ground-truth test data and a perturbed dataset, always matching the set sizes to have the same $n$. The three perturbations considered are a shift, scale and correlation change, applied to one of the two datasets at a time. Shift and scale are applied to only one parameter (out of the 13) at a time. The change in correlation only affects one distinct parameter, conditioned on one other parameter. Concretely, for the shift perturbation, all values of the dataset are increased by the discrepancy scale $\Delta$ we set, i.e. $x_j \rightarrow x_j+\Delta\sigma_j$. Here, $\sigma_j$ refers to the parameter's standard deviation. Similarly, a change in scale widens a parameter distribution as $x_j \rightarrow \mu_j + (1+\Delta)(x_j-\mu_j)$, where $\mu_j$ is the distribution's mean. Lastly, we include a correlation coefficient that scales as $\approx \Delta$, i.e. $x_j \rightarrow x_j + \Delta\sigma_jz_l$, with $z_l = (x_l-\mu_l)/\sigma_l$ and $l\neq j$. A realistic NF failure would likely distort several coordinates at once, producing a larger mismatch effect than a single-column discrepancy. Therefore, adding the perturbation to one parameter only likely results in a conservative estimate of the required number of test samples. Because we use the true injection data, the test inherits all of its correlations, skew, boundaries, etc.

Two rates characterize the decision of whether two sets of samples come from the same distribution. The false-positive rate $\alpha$ quantifies how often a \emph{correct} NF is rejected, and we use $\alpha = 0.05$. ``Correct'' in this case means that when sampling from the NF any discrepancy in shift, scale, or correlation is smaller than $\Delta$. The power $1-\beta$ describes the rate at which a \emph{wrong} NF is correctly rejected. Power depends on how wrong the NF is, how many samples are available, and the test itself. Writing $\pi(\Delta,n)$ for the probability of rejection at discrepancy $\Delta$ and $n$ samples per set, we have $\pi(\Delta = 0,n) = \alpha$ by construction, and we require

\begin{align} \label{eq:TestPower}
    \pi(\Delta, n) \geq 1-\beta \ .
\end{align}

The three quantities $\alpha$, $\Delta$, and $1-\beta$ therefore fix $n$ \cite{Cohen2013}. However, $n$ is not universal across tests; instead, there is one $n$ per test. We fix $\alpha=0.05$ and $1-\beta=0.95$ and aim to resolve a $\Delta = 0.1\sigma$. We then solve for $n$.

For the KS and AD tests, we brute-force the power curve and estimate $n$. After perturbing one of the datasets, we run the test many times for some candidate $n$ and obtain the power estimate as $\hat{\pi}(n) = N_\mathrm{rejections}/N_\mathrm{repetitions}$. We use $N_\mathrm{repetitions}=400$. Because $\pi$ is monotone in $n$, the smallest $n$ that fulfills $\hat{\pi}(n) \geq 1-\beta$ is found numerically by bracketing and bisecting $n$ until the threshold is reached.

The brute-force estimate for the PQMass test and the C2ST is computationally expensive. We avoid this by using analytic approximations. For the C2ST, the test statistic is the AUC of the trained classifier, which is asymptotically normal. The power curve follows analytically once the AUC's mean and variance are known. Following \cite{Obuchowski1997}, the required sample size satisfies

\begin{align}
    n = \frac{\bigl(z_{1-\alpha/2}\sqrt{V_0(n)} + z_{1-\beta}\sqrt{V_1(n)}\bigr)^{2}}
           {\left(\mathrm{AUC} - \tfrac{1}{2}\right)^{2}},
  \label{eq:c2st_n}
\end{align}

where $\mathrm{AUC}$ is the AUC achieved by the trained classifier, $1/2$ is its value under the null, $V_0$ and $V_1$ are the variances of the AUC estimator under the null and the alternative, and $z_q$ denotes the $q$-quantile of the standard normal distribution. Both variances are computed following \cite{Hanley1982} and \cite{DeLong1988}. Note that due to the training and validation split of the data for the classifier, $n$ should be appropriately rescaled for the C2ST. 

Unlike the C2ST, the PQMass test has a known null distribution of the test statistic. Its statistic is a Pearson $\chi^2$ over $k+1$ Voronoi cells, so under the null hypothesis it follows a $\chi^2_k$ distribution. We therefore reject the null hypothesis at level $\alpha$ when $X^2 > \chi^2_{k,1-\alpha}$. The alternative statistic follows a non-central $\chi^{2}_k(\lambda)$ \cite{Mitra1958}. Thus, under the alternative hypothesis,
\begin{align}
    X^2 \sim \chi^{2}_k(\lambda),
\end{align}
where $\lambda$ controls how strongly the alternative shifts the distribution of the test statistic $X^2$ towards larger values. We express $\lambda$ in terms of the expected difference between the fractions of samples that fall into each Voronoi cell. Let $\hat{p}_m$ and $\hat{p}^{(\Delta)}_m$ denote the fractions of samples from the two datasets falling into cell $m$, where the superscript $^{(\Delta)}$ refers to the perturbed dataset. The statistic for equal dataset sizes then follows as $X^{2} = \frac{n}{2}\sum_m (\hat p_m - \hat p^{(\Delta)}_m)^{2}/\bar p_m$, where $\bar p_m=(\hat p_m + \hat p^{(\Delta)}_m)/2$ is the pooled probability to fall into cell $m$ given the null and alternative pair. The summation term is expected to depend only on the two underlying distributions (null and alternative) and the Voronoi partition, but not on $n$. We can therefore separate the sample-size dependence from the distributional difference by replacing the observed differences with their expectations, $\delta_m = \mathbb{E}[\hat p_m - \hat p^{(\Delta)}_m]$. We define:

\begin{align}\label{eq:pqm_lambda}
  \lambda = \frac{n}{2}\,W, \qquad W = \sum_m \frac{\delta_m^{2}}{\bar p_m}
\end{align}

We estimate $W$ once for a given discrepancy $\Delta$ from the full injection set and average it over random tessellations. The required sample size then follows by solving

\begin{align}\label{eq:pqm_power}
  1-\beta = P_\lambda\left(X^2 > \chi^{2}_{k,1-\alpha}\right) = 1 - F_{\chi^{2}_k(\lambda)}\!\left(\chi^{2}_{k,1-\alpha}\right),
\end{align}

which is monotone in $n$ and solved numerically. $F$ denotes the CDF, thus we compute the probability that the non-central $\chi^2$ statistic falls above the critical value given by $\alpha$. Because $\delta_m = \mathcal{O}(\Delta)$, we have $W \propto \Delta^{2}$ and hence $n \propto \Delta^{-2}$: resolving half the mismatch costs four times as many samples.

\section{Additional Evaluation Details\label{sec:appendixC_additionalvalidation}} 

\subsection{Comparison of Summary Statistics\label{subsec:absRelDifferenceSSTandCorr}} 

\begin{figure*}[t]
    \centering

    \begin{subfigure}[t]{0.48\textwidth}
        \centering
        \includegraphics[width=\linewidth]{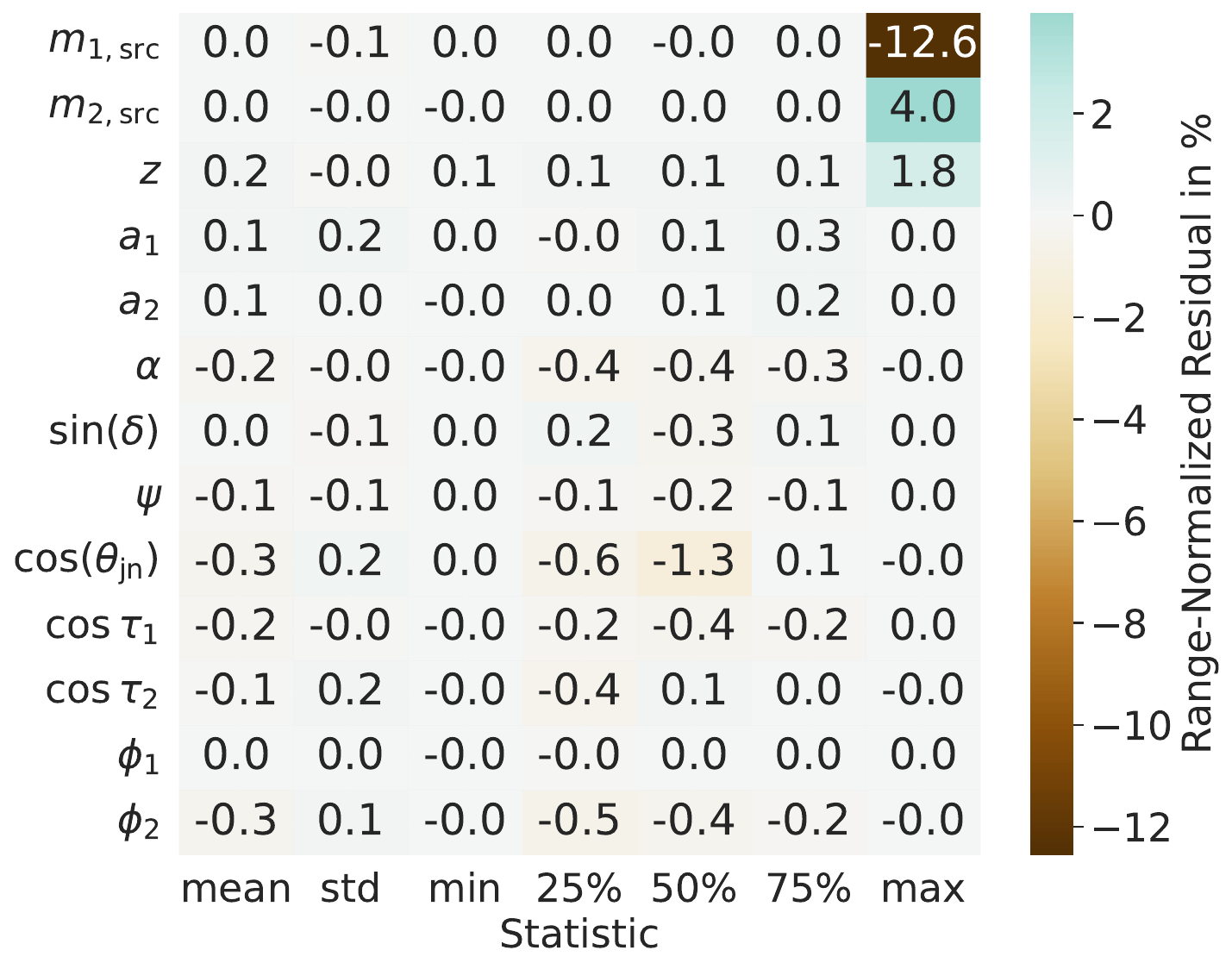}
        \caption{Residual of NF samples relative to the test set, in \%, normalized over the range of each parameter in the test set. The largest differences can be observed for the maximum values of the masses and redshift. These three parameters are unbounded in reality and therefore no upper limit is imposed on the NF.}
        \label{fig:summaryquantiles_NFsamples}
    \end{subfigure}
    \hfill
    \begin{subfigure}[t]{0.48\textwidth}
        \centering
        \includegraphics[width=\linewidth]{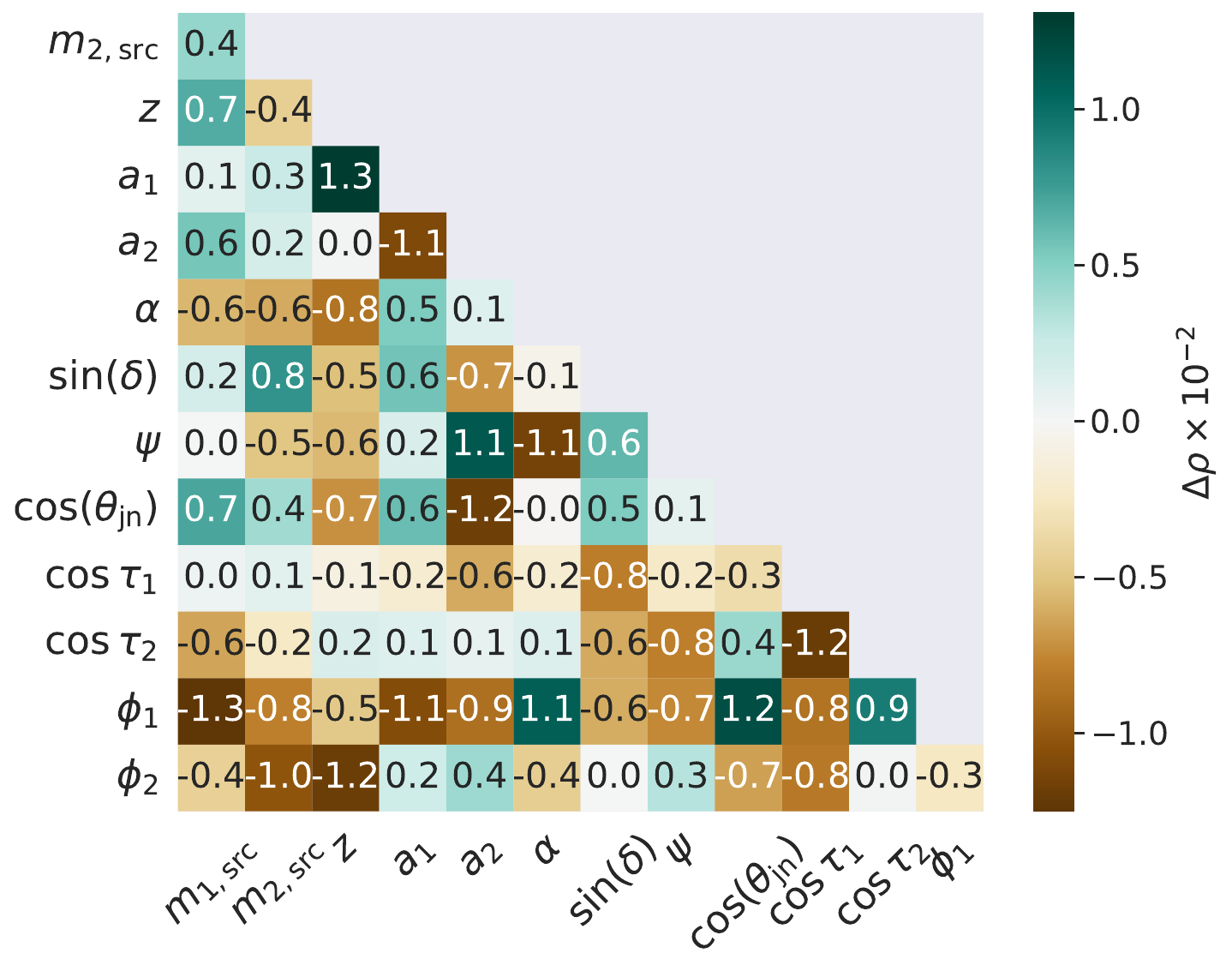}
        \caption{Difference in the correlation matrices between the test set and NF samples.
        For each dataset, we compute the Pearson correlation matrix and subtract the test set correlations from the NF samples' correlations. All differences in the correlation coefficients are below 0.014 and therefore quite small.}
        \label{fig:correlations_NFsamples}
    \end{subfigure}

    \caption{Comparison of NF samples with test set summary statistics and parameter correlations.}
    \label{fig:summary_statistics}
\end{figure*}

In the following, we compare the summary statistics (minimum, mean and maximum values, as well as the 25\%, 50\%, and 75\% quantiles) for the NF samples and the test set. Generally, we find good agreement between all sets. We show the comparison of the NF samples with the test set in Figures \ref{fig:summaryquantiles_NFsamples} and \ref{fig:correlations_NFsamples}. The comparison of the emulator detections against the test set yields very similar results, and therefore we omit them here. The only noticeable difference is observed in the maximum values of the primary and secondary masses. 

\subsection{Joint and Marginal parameter distributions of detected events\label{subsec:jointAndMarginal13d}} 

\begin{figure*}[t]
    \centering
    \includegraphics[width=1\textwidth]{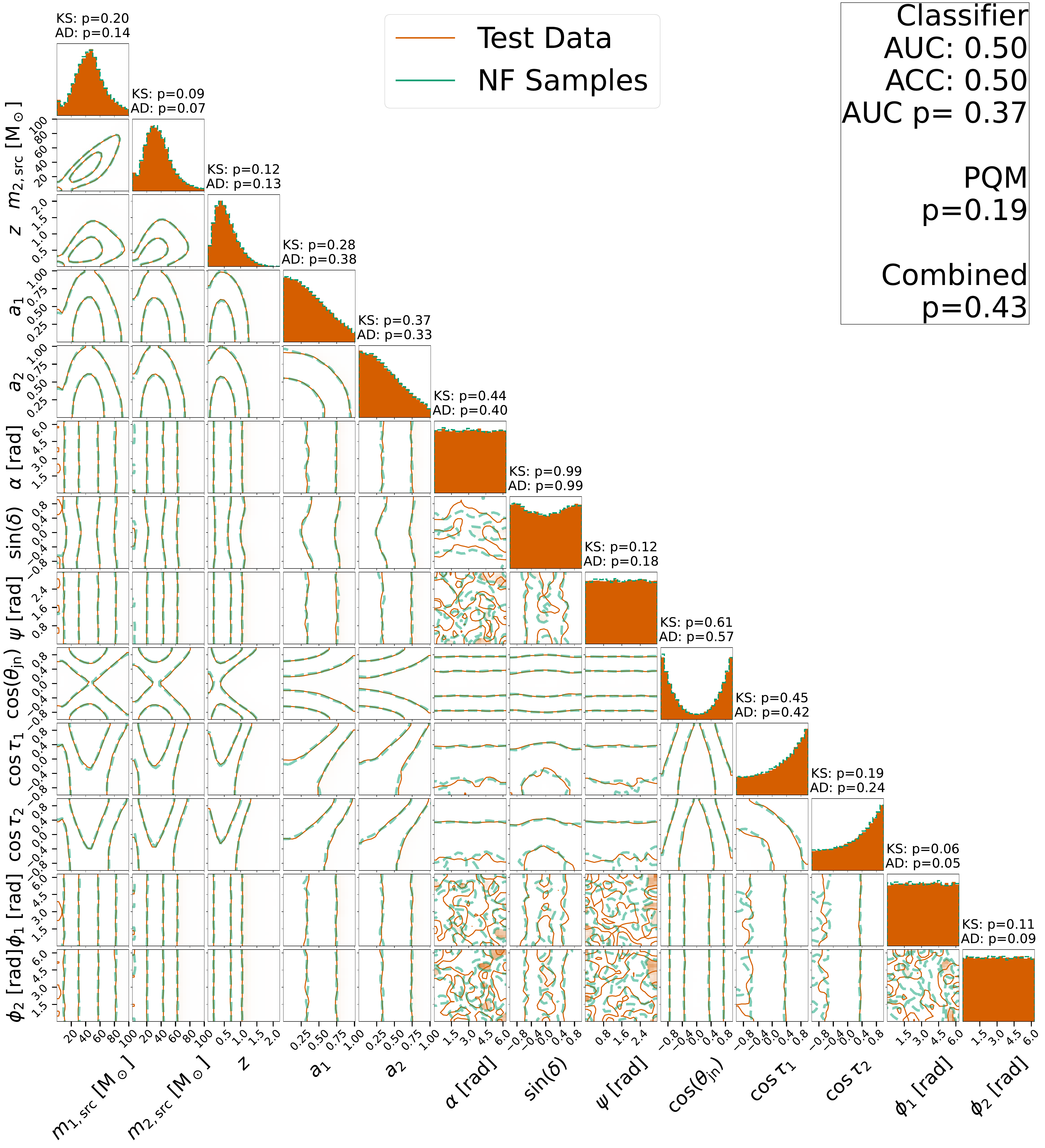}
    \caption{Joint and marginal parameter distributions of detected events for all of the 13 parameters we account for. Green displays the data sampled from the NF, whereas orange depicts the ground-truth test data. All parameter distributions are well reproduced as indicated by the metrics. }
    \label{fig:appC_jointAndMarginal13dSampled}
\end{figure*}

One central part of the evaluation of our proposed method is that the parameter distributions of detected events in the injection data are correctly reproduced, both when sampling from the NF and when generating emulator detections via the procedure outlined in section \ref{subsec:nffoundeval}. We therefore display the full 13-parameter joint and marginal distributions here, showing the correct reproduction of the parameter distributions in both cases. We begin with the joint and marginal distributions for the NF samples in Figure \ref{fig:appC_jointAndMarginal13dSampled}, overlaid on the ground-truth test data. 
The KS, AD, and PQMass tests as well as the classifier metrics quantitatively confirm the visual inspection. This confirms that we can reproduce the correct parameter distribution of detected events by sampling from the NF.

\begin{figure*}[t]
    \centering
    \includegraphics[width=1\textwidth]{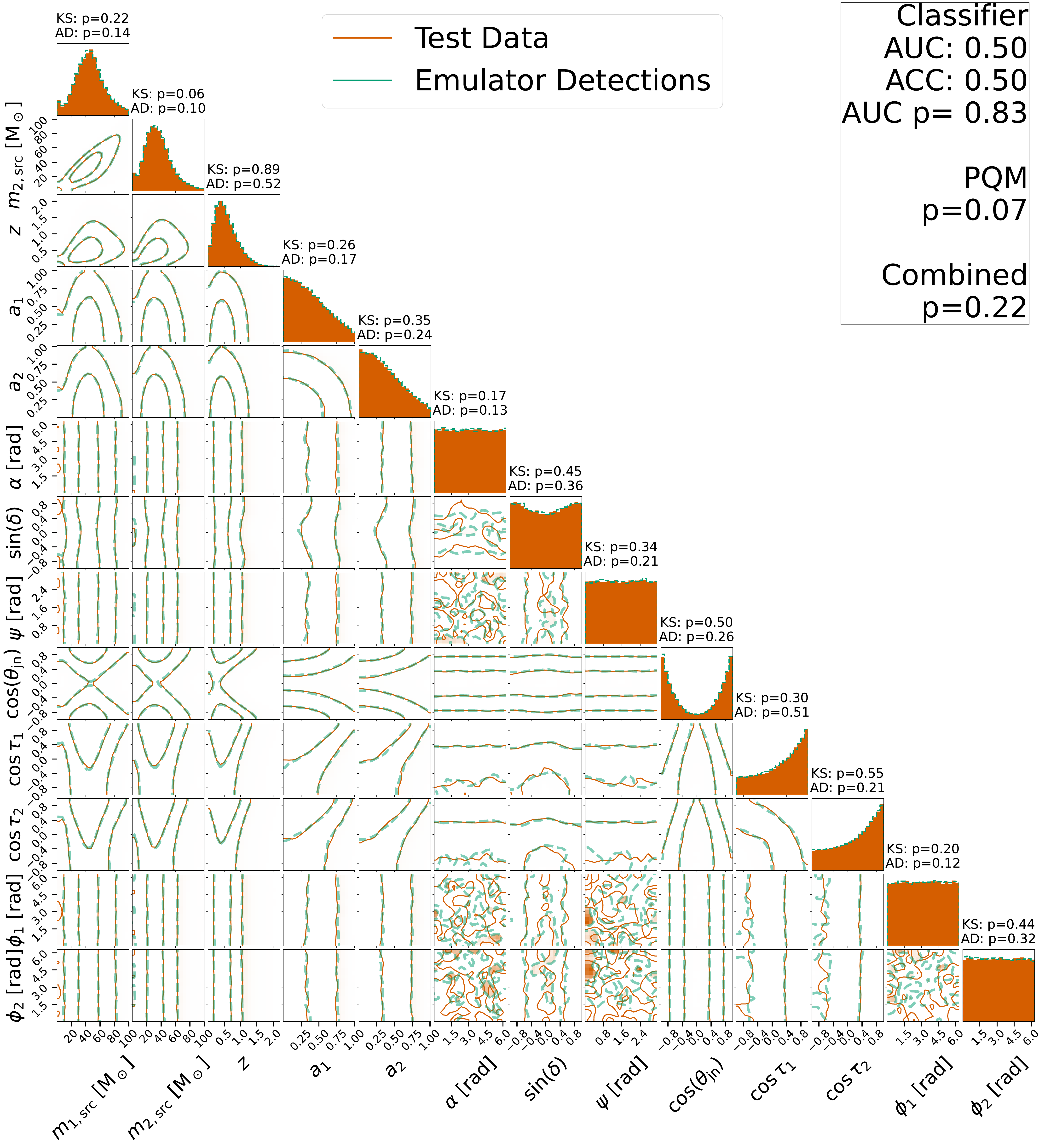}
    \caption{Joint and marginal parameter distributions of detected events for all of the 13 parameters we account for. Green displays the data probabilistically labeled as detected using the end-to-end NF emulator, whereas orange depicts the ground-truth test data. All parameter distributions are well reproduced as indicated by the metrics.}
    \label{fig:appC_jointAndMarginal13dFound}
\end{figure*}

In Figure \ref{fig:appC_jointAndMarginal13dFound} we show the joint and marginal distributions of the test set, but now overlay the parameter distributions of the emulator detections, as described in section \ref{subsec:nffoundeval}. Instead of drawing samples directly from the NF as we did in section \ref{subsec:nfstandaloneeval}, we draw binary merger parameters $\theta_e$ from the injection population model, evaluate detectability with the NF emulator and probabilistically accept events as detected. Again, we barely see the underlying test data in most cases, because the parameter distributions are very well reproduced.

We find a very low number of over-predictions in a plausible parameter region very similar to that in section \ref{subsec:plausibilitychecks}. Further, the over-predictions occur rarely, where most over-predictions exceed the 100\% limit by a small factor of 1.0-1.2.

\clearpage
\bibliography{bib}

\end{document}